\documentclass[12pt]{article}

\input epsf
\usepackage{amssymb,amsmath,wrapfig,subfigure}
\usepackage[matrix,arrow,curve]{xy}
\usepackage{cite}
\usepackage{graphicx,color}
\usepackage[debug,pageanchor=false]{hyperref}
\definecolor{link}{rgb}{.8,.15,.1}
\hypersetup{colorlinks=true,linkcolor=link,citecolor=link,urlcolor=link,linktocpage}
\usepackage{cancel}

\makeatletter
\@addtoreset{equation}{section}
\makeatother

\def\pp {{\mathbb{P}}}
\def\rr {{\mathbb{R}}}
\def\zz {{\mathbb{Z}}}
\def\del {\partial}

\def\del {\partial}

\def\vol {\mathrm{vol}}

\def\lsim{\mathrel{\rlap{\lower4pt\hbox{\hskip1pt$\sim$}}
    \raise1pt\hbox{$<$}}}                

\begin{document}

\begin{titlepage}

\begin{center}

\vskip .3in \noindent

{\Large \bf{All AdS$_7$ solutions of type II supergravity}}

\bigskip

	Fabio Apruzzi$^1$, Marco Fazzi$^2$, Dario Rosa$^3$ and Alessandro Tomasiello$^3$\\

       \bigskip
	 {\small $^1$ Institut f\"ur Theoretische Physik, Leibniz Universit\"at Hannover,
	Appelstra\ss e 2, 30167 Hannover, Germany \\
	 \vspace{.1cm}
	 $^2$ Universit\'e Libre de Bruxelles and International Solvay Institutes, ULB-Campus Plaine CP231, B-1050 Brussels, Belgium \\
	\vspace{.1cm} 
	$^3$ Dipartimento di Fisica, Universit\`a di Milano--Bicocca, I-20126 Milano, Italy\\
    and\\
    INFN, sezione di Milano--Bicocca
	}

       \vskip .5in
       {\bf Abstract }
       \vskip .1in
\end{center}

	In M-theory, the only AdS$_7$ supersymmetric solutions are AdS$_7\times S^4$ and its orbifolds. In this paper, we find and classify new supersymmetric solutions of the type AdS$_7 \times M_3$ in type II supergravity. While in IIB none exist, in IIA with Romans mass (which does not lift to M-theory) there are many new ones. We use a pure spinor approach reminiscent of generalized complex geometry. Without the need for any Ansatz, the system determines uniquely the form of the metric and fluxes, up to solving a system of ODEs. Namely, the metric on $M_3$ is that of an $S^2$ fibered over an interval; this is consistent with the Sp(1) R-symmetry of the holographically dual (1,0) theory. By including D8 brane sources, one can numerically obtain regular solutions, where topologically $M_3 \cong S^3$.

\noindent

\vfill
\eject

\end{titlepage}

\hypersetup{pageanchor=true}

\tableofcontents

\section{Introduction} 
\label{sec:intro}

Interacting quantum field theories generally become hard to define in more than four dimensions. A Yang--Mills theory, for example, becomes strongly coupled in the UV. In six dimensions, a possible alternative would be to use a two-form gauge field. Its nonabelian formulation is still unclear, but  string theory predicts that a $(2,0)$-superconformal completion of such a field actually exists on the worldvolume of M5-branes. Understanding these branes is still one of string theory's most interesting challenges.

This prompts the question of whether other non-trivial six-dimensional theories exist. There are in fact several other string theory constructions \cite{blum-intriligator,brunner-karch,hanany-zaffaroni-6d,ferrara-kehagias-partouche-zaffaroni} that would engineer such theories. Progress has also been made (see for example \cite{samtleben-sezgin-wimmer,chu}) in writing explicitly their classical actions. 

Another way to establish the existence of superconformal theories in six dimensions is to look for supersymmetric AdS$_7$ solutions in string theory. In this paper, we classify such solutions. As we will review later, in M-theory, one only has AdS$_7\times S^4$ (which is holographically dual to the $(2,0)$ theory) or an orbifold thereof. That leaves us with AdS$_7 \times M_3$ in IIA with non-zero Romans mass $F_0\neq 0$ (which cannot be lifted to M-theory) or in IIB. 

Here we will show that, while there are no such solutions in IIB, many do exist in IIA with non-zero Romans mass $F_0$.

Our methods are reminiscent of the generalized complex approach for 
Mink$_4\times M_6$ or AdS$_4\times M_6$ solutions \cite{gmpt2}. We start with a similar system \cite{lust-patalong-tsimpis} for Mink$_6\times M_4$, and we then use the often-used trick of viewing AdS$_7$ as a warped product of Mink$_6$ with a line. This allows us to obtain a system valid for AdS$_7\times M_3$. A similar procedure was applied in \cite{gabella-gauntlett-palti-sparks-waldram} to derive a system for AdS$_5 \times M_5$ from Mink$_4 \times M_6$. The system we derive is written in terms of differential forms satisfying some algebraic constraints; mathematically, these constraints mean that the forms define a generalized identity$\times$identity structure on $T_{M_3} \oplus T^*_{M_3}$. This fancy language, however, will not be needed here; we will give a parameterization of such structures in terms of a vielbein $\{e_a\}$ and some angles, and boil the system down to one written in terms of those quantities.

When one writes supersymmetry as a set of PDEs in terms of forms, they may have some interesting geometrical interpretation (such as the one in terms of generalized complex geometry in \cite{gmpt2}); but, to obtain solutions, one usually needs to make some Ansatz, such as that the space is homogeneous or that it has cohomogeneity one. One then reduces the differential equations to algebraic equations or to ODEs, respectively. 

The AdS$_7 \times M_3$ case is different. As we will see, the equations actually determine explicitly the vielbein $\{ e_a \}$ in terms of derivatives of our parameterization function. This gives a local, explicit form for the metric, without any Ansatz. By a suitable redefinition we find that the metric describes an $S^2$ fibration over a one-dimensional space. 

This is actually to be expected holographically. A $(1,0)$ superconformal theory has an Sp(1)$\cong$SU(2) R-symmetry group, which should appear as the isometry group of the internal space $M_3$. With a little more work, all the fluxes can also be determined, and they are also left invariant by the SU(2) isometry group of our $S^2$ fiber. All the Bianchi identities and equations of motion are automatically satisfied, and existence of a solution is then reduced to a system of two coupled ODEs.\footnote{This is morally a hyper-analogue to the reduction performed in \cite{gabella-gauntlett-palti-sparks-waldram} along the generalized Reeb vector, although in our case the situation is so simple that we need not introduce that reduction formalism.} From this point on, our analysis is pretty standard: in order for $M_3$ to be compact, the coordinate $r$ on which everything depends should in fact parameterize an interval $[r_{\rm N}, r_{\rm S}]$, and the $S^2$ should shrink at the two endpoints of the interval, which we from now on will call ``poles''. This requirement translates into certain boundary conditions for the system of ODEs.

We have studied the system numerically. We can obtain regular\footnote{On the loci where branes are present, the metric is of course not regular, but such singularities are as usual excused by the fact that we know that D-branes have an alternative definition as boundary conditions for open strings, and are thought to be objects in the full theory. The singularity is particularly mild for D8's, which manifest themselves as jumps in the derivatives of the metric and other fields --- which are themselves continuous.} solutions if we insert brane sources. We exhibit solutions with D6's, and solutions with one or two D8 stacks, appropriately stabilized by flux. For example, in the solution with two D8 stacks,  they have opposite D6 charge, and their mutual electric attraction is balanced against their gravitational tendency to shrink.  (For D8-branes, there is no problem with the total D-brane charge in a compact space; usually such problems are found by integrating the flux sourced by the brane over a sphere surrounding the brane, whereas for a D8 such a transverse sphere is simply an $S^0$.) We think that there should exist generalizations with an arbitrary number of stacks.

It is natural to think that our regular solutions with D8-branes might be related to D-brane configurations in \cite{brunner-karch,hanany-zaffaroni-6d}, which should indeed engineer six-dimensional $(1,0)$ superconformal theories. Supersymmetric solutions for configurations of that type have actually been found \cite{janssen-meessen-ortin} (see also \cite{imamura-D8}); non-trivially, they are fully localized. It is in principle possible that their results are related to ours by some limit. Such a relationship is not obvious, however, in part because of the SU(2) symmetry, that forces our sources to be only parallel to the $S^2$-fiber. It would be interesting to explore this possibility further.

We will begin our analysis in section \ref{sec:pure} by finding the pure spinor system (\ref{eq:73}) relevant for supersymmetric AdS$_7\times M_3$ solutions. In section \ref{sec:para} we will then derive the parameterization (\ref{eq:psigen}) for the pure spinors in terms of a vielbein and some functions. In section \ref{sec:gen} will then use this parameterization to analyze the system (\ref{eq:73}). As we mentioned, we will reduce the problem to a system of ODEs; regularity imposes certain boundary conditions on this system. Fluxes and metric are fully determined by a solution to the system of ODEs. Finally, in section \ref{sec:exp} we study the system numerically, finding some regular examples, shown in figures \ref{fig:1d8} and \ref{fig:2d8}.


\section{Supersymmetry and pure spinor equations in three dimensions} 
\label{sec:pure}

In this section, we will derive a system of differential equations on forms in three dimensions that is equivalent to preserved supersymmetry for solutions of the type AdS$_7\times M_3$. We will derive it by a commonly-used trick: namely, by considering AdS$_{d+1}$ as a warped product of Mink$_d$ and $\rr$. We will begin in section \ref{sub:64} by reviewing a system equivalent to supersymmetry for Mink$_6 \times M_4$. In section \ref{sub:73} we will then translate it to a system for AdS$_7\times M_3$.

\subsection{${\rm Mink}_6\times M_4$} 
\label{sub:64}

Preserved supersymmetry for Mink$_4\times M_6$ was found \cite{gmpt2} to be equivalent to the existence on $M_6$ of an ${\rm SU}(3) \times {\rm SU}(3)$ structure satisfying certain differential equations reminiscent of generalized complex geometry \cite{hitchin-gcy,gualtieri}.

Similar methods can be useful in other dimensions. For Mink$_6\times M_4$ solutions, \cite{lust-patalong-tsimpis} found a system in terms of ${\rm SU}(2) \times {\rm SU}(2)$ structure on $M_4$, described by a pair of pure spinors $\phi^{1,2}$. Similarly to the Mink$_4\times M_6$ case, they can be characterized in two ways. One is as bilinears of the internal parts $\eta^{1,2}$ of the supersymmetry parameters in (\ref{eq:eps64}):\footnote{As usual, we are identifying forms with bispinors via the Clifford map $dx^{m_1}\wedge \ldots \wedge dx^{m_k}\mapsto \gamma^{m_1\ldots m_k}$. ${}_\mp$ denotes chirality, and $\eta^c \equiv B_4 \eta^*$ denotes Majorana conjugation; for more details see appendix \ref{app:gamma}. The factors $e^{-A_4}$ are included for later convenience.}
\begin{equation}\label{eq:phi12}
	\phi^1_\mp= e^{-A_4}\eta^1_+ \otimes \eta_\mp^{2\,\dagger}\ ,\qquad 
	\phi^2_\mp= e^{-A_4}\eta^1_+ \otimes \eta_\mp^{2c\,\dagger}\ ,
\end{equation}
where the warping function $A_4$ is defined by 
\begin{equation}\label{eq:A4}
	ds^2_{10}= e^{2A_4} ds^2_{{\rm Mink}_6} + ds^2_{M_4}\ .
\end{equation}
The upper index in (\ref{eq:phi12}) is relevant to IIA, the lower index to IIB; so in IIA we have that $\phi^{1,2}$ are both odd forms, and in IIB that they are both even. One can also give an alternative characterization of $\phi^{1,2}$, as a pair of pure spinors which are \emph{compatible}. This stems directly from their definition as an ${\rm SU}(2) \times {\rm SU}(2)$ structure, and it means that the corresponding generalized almost complex structures commute. This latter constraint can also be formulated purely in terms of pure spinors as $(\phi^1,\phi^2)=(\bar \phi^1, \phi^2)$.\footnote{As usual, the Chevalley pairing in this equation is defined as $(\alpha,\beta)= (\alpha \wedge \lambda(\beta))_{\rm top}$; $\lambda$ is the sign operator defined on $k$-forms as $\lambda \omega_k \equiv (-)^{\lfloor \frac k2 \rfloor} \omega_k$.} This can be shown similarly to an analogous statement in six dimensions; see \cite[App.~A]{t-reform}.
 
The system equivalent to supersymmetry now reads \cite{lust-patalong-tsimpis}
\footnote{We have massaged a bit the original system in \cite{lust-patalong-tsimpis}, by eliminating ${\rm Re} \phi^1_\mp$ from the first equation of their (4.11).}
\begin{subequations}\label{eq:64}
\begin{align}
	& d_H \bigl(e^{2A_4-\phi} \mathrm{Re} \phi^1_{\mp}\bigr) = 0 \ ,\label{eq:64R1} \\
   & d_H \bigl(e^{4A_4-\phi} \mathrm{Im} \phi^1_{\mp}\bigr) = 0 \label{eq:64I1}\ , \\
   & d_H \bigl(e^{4A_4-\phi} \phi^2_{\mp}\bigr) = 0 \label{eq:642} \ ,\\
	& e^\phi F = \mp 16 *_4 \lambda (dA_4 \wedge {\rm Re} \phi^1_\mp)\ , \\
	& (\overline{\phi^1_\pm},\phi^1_\pm)=(\overline{\phi^2_\pm},\phi^2_\pm)= \frac14\ . \label{eq:64norm}
\end{align}
\end{subequations}
Here, $\phi$ is the dilaton; $d_H=d-H\wedge$ is the twisted exterior derivative; $A_4$ was defined in (\ref{eq:A4});
$F$ is the internal RR flux, which, as usual, determines the external flux via self-duality:
\begin{equation}
	F_{(10)} \equiv F + e^{6A_4} \mathrm{vol}_6 \wedge \ast_4 \lambda F\ .
\end{equation}

Actually, (\ref{eq:64}) contains an assumption: that the norms of the $\eta^i$ are equal. For a noncompact $M_4$, it might be possible to have different norms; (\ref{eq:64}) would then have to be slightly changed. (See \cite[Sec.~A.3]{gmpt3} for comments on this in the Mink$_4\times M_6$ case.) As shown in appendix \ref{app:gamma}, however, for our purposes such a generalization is not relevant.

With this caveat, the system (\ref{eq:64}) is equivalent to supersymmetry for Mink$_6\times M_4$. It can be found by direct computation, or also as a consequence of the system for Mink$_4\times M_6$ in \cite{gmpt2}: one takes $M_6=\rr^2\times M_4$, with warping $A=A_4$, internal metric $ds^2_{M_6}=e^{2A_4}((dx^4)^2+ (dx^5)^2) + ds^2_{M_4}$, and, in the language of \cite{gmpt3},
\begin{equation}
	\Phi_1 = e^{A_4} (dx^4+idx^5)\wedge \phi^2_\mp \ ,\qquad \Phi_2 = (1+ i e^{2A_4} dx^4\wedge dx^5)\wedge \phi^1_\mp\ .
\end{equation}
Furthermore, (\ref{eq:64}) can also be found as a consequence of the ten-dimensional system in \cite{10d}. \cite{lust-patalong-tsimpis} also give an interpretation of the system in terms of calibrations, along the lines of \cite{martucci-smyth}. 


\subsection{${\rm AdS}_7\times M_3$} 
\label{sub:73}

As we anticipated, we will now use the fact that AdS can be used as a warped product of Minkowski space with a line. We would like to classify solutions of the type AdS$_7\times M_3$. These in general will have a metric 
\begin{equation}\label{eq:met73}
	ds^2_{10} = e^{2 A_3} ds^2_{{\rm AdS}_7} + ds^2_{M_3}
\end{equation}
where $A_3$ is a new warping function (different from the $A_4$ in (\ref{eq:A4})). Since
\begin{equation}\label{eq:ads7}
	 ds^2_{{\rm AdS}_7} = \frac{d \rho^2}{\rho^2} + \rho^2 ds^2_{{\rm Mink}_6}\ ,
\end{equation}
(\ref{eq:met73}) can be put in the form (\ref{eq:A4}) if we take
\begin{equation}\label{eq:43}
	e^{A_4} = \rho e^{A_3} \ ,\qquad ds^2_{M_4} = \frac{e^{2A_3}}{\rho^2}d \rho^2 + ds^2_{M_3}\ .
\end{equation}
A genuine AdS$_7$ solution is one where not only the metric is of the form (\ref{eq:ads7}), but where there are also no fields that break its SO(6,2) invariance. This can be easily achieved by additional assumptions: for example, $A_3$ should be a function of $M_3$. The fluxes $F$ and $H$, which in section \ref{sub:64} were arbitrary forms on $M_4$, should now be forms on $M_3$. For IIA, $F=F_0 + F_2 + F_4$: in order not to break SO$(6,2)$, we impose $F_4=0$, since it would necessarily have a leg along AdS$_7$; for IIB, $F=F_1+F_3$.

Following this logic, solutions to type II equations of motion of the form AdS$_7\times M_3$ are a subclass of solutions of the form Mink$_6\times M_4$. In appendix \ref{app:gamma}, we also show how the AdS$_7\times M_3$ supercharges get translated in the Mink$_6\times M_4$ framework, and that the internal spinors have equal norm, as we anticipated in section \ref{sub:64}. Using (\ref{eq:etachi}), we also learn how to express the $\phi^{1,2}$ in (\ref{eq:phi12}) in terms of bilinears of spinors $\chi_{1,2}$ on $M_3$:
\begin{equation}\label{eq:phipsi}
	\phi^1_\mp = \frac12 \left(\psi^1_\mp + i e^{A_3} \frac{d \rho}{\rho} \wedge \psi^1_\pm\right) \ ,\qquad
	\phi^2_\mp = \mp \frac12 \left(\psi^2_\mp + i e^{A_3} \frac{d \rho}{\rho} \wedge \psi^2_\pm\right)\ ,
\end{equation}
with
\begin{equation}\label{eq:psichi}
	\psi^1= \chi_1 \otimes \chi_2^\dagger \ ,\qquad \psi^2 = \chi_1 \otimes \chi_2^{c\,\dagger}\ .
\end{equation}
As in section \ref{sub:64}, we have implicitly mapped forms to bispinors via the Clifford map, and in (\ref{eq:phipsi}) the subscripts $_\pm$ refer to taking the even or odd form part. (Recall also that $\phi^{1,2}_-$ is relevant to IIA, and $\phi^{1,2}_+$ to IIB; see (\ref{eq:64}).) The spinors $\chi_{1,2}$ have been taken to have unit norm.

$\psi^{1,2}$ are differential forms on $M_3$, but not just any forms. (\ref{eq:psichi}) imply that they should obey some algebraic constraints. Those constraints could be interpreted in a fancy way as saying that they define an identity$\times$identity structure on $T_{M_3} \oplus T^*_{M_3}$. However, three-dimensional spinorial geometry is simple enough that we can avoid such language: rather, in section \ref{sec:para} we will give a parameterization that will allow us to solve all the algebraic constraints resulting from (\ref{eq:psichi}).

We can now use (\ref{eq:phipsi}) in (\ref{eq:64}). Each of those equations can now be decomposed in a part that contains $d\rho$ and one that does not. Thus, the number of equations would double. However, for (\ref{eq:64R1}), (\ref{eq:64I1}) and (\ref{eq:642}), the part that does not contain $d\rho$ actually follows from the part that does. The ``norm'' equation, (\ref{eq:64norm}), simply reduces to a similar equation for a three-dimensional norm. Summing up: 
\begin{subequations}\label{eq:73}
\begin{align}
	&d_H {\rm Im} (e^{3A_3-\phi} \psi^1_\pm) = -2 e^{2A_3-\phi} {\rm Re} \psi^1_\mp\ , 
	\label{eq:73I1}\\
	&d_H {\rm Re} (e^{5A_3-\phi} \psi^1_\pm) = 4 e^{4A_3-\phi} {\rm Im} \psi^1_\mp\ ,
	\label{eq:73R1}\\
	&d_H (e^{5A_3-\phi} \psi^2_\pm) = -4i e^{4A_3-\phi} \psi^2_\mp\ ,
	\label{eq:732}\\
	&\pm \frac18 e^\phi *_3 \lambda F = dA_3 \wedge {\rm Im} \psi^1_\pm + e^{-A_3} {\rm Re} \psi^1_\mp\ , \label{eq:73f}\\ 
	&dA_3 \wedge {\rm Re} \psi^1_\mp = 0 \label{eq:73dAR}\ , \\
	&(\psi^{1,2}_+,\overline{\psi^{1,2}_-})=-\frac i2  \label{eq:73norm}\ ;
\end{align}
\end{subequations} 
again with the upper sign for IIA, and the lower for IIB.

The system (\ref{eq:73}) is equivalent to supersymmetry for AdS$_7 \times M_3$. As we show in appendix \ref{app:gamma}, a supersymmetric AdS$_7 \times M_3$ solution can be viewed as a supersymmetric Mink$_6\times M_4$ solution, and for this the system (\ref{eq:64}) is equivalent to supersymmetry. (\ref{eq:73}) can also be obtained directly from the ten-dimensional system in \cite{10d}, but other equations also appear, and extra work is needed to show that those extra equations are redundant. 

In (\ref{eq:73}) the cosmological constant of AdS$_7$ does not appear directly, since we have taken its radius to be one in (\ref{eq:ads7}). We did so because a non-unit radius can be reabsorbed in the factor $e^{2 A_3}$ in (\ref{eq:met73}). 

Before we can solve (\ref{eq:73}), we have to solve the algebraic constraints that follow from the definition of $\psi^{1,2}$ in (\ref{eq:psichi}); we will now turn to this problem.



\section{Parameterization of the pure spinors} 
\label{sec:para}

In section \ref{sub:73} we obtained a system of differential equations, (\ref{eq:73}), that is equivalent to supersymmetry for an AdS$_7\times M_3$ solution. The $\psi^{1,2}$ appearing in that system are not arbitrary forms; they should have the property that they can be rewritten as bispinors (via the Clifford map $dx^{i_1} \wedge \ldots \wedge dx^{i_k} \mapsto \gamma^{i_1\ldots i_k}$) as in (\ref{eq:psichi}). In this section, we will obtain a parameterization for the most general set of $\psi^{1,2}$ that has this property. This will allow us to analyze (\ref{eq:73}) more explicitly in section \ref{sec:gen}. 

We will begin in section \ref{sub:one} with a quick review of the case $\chi_1=\chi_2$, and then show in section \ref{sub:two} how to attack the more general situation where $\chi_1\neq \chi_2$. 

\subsection{One spinor} 
\label{sub:one}

We will use the Pauli matrices $\sigma_i$ as gamma matrices, and use $B_3=\sigma_2$ as a conjugation matrix (so that $B_3 \sigma_i = - \sigma_i^t B_3 = - \sigma_i^* B_3$). We will define
\begin{equation}
	\chi^c \equiv B_3 \chi^* \ ,\qquad \overline{\chi}\equiv\chi^t B_3\ ;
\end{equation}
notice that $\chi^{c\,\dagger}= \chi^t B_3^\dagger = \overline{\chi}$. 

We will now evaluate $\psi^{1,2}$ in (\ref{eq:psichi}) when $\chi_1=\chi_2 \equiv \chi$; as we noted in section \ref{sub:73}, $\chi$ is normalized to one. Notice first a general point about the Clifford map $\alpha_k = \frac1{k!} \alpha_{i_1\ldots i_k}dx^{i_1}\wedge\ldots \wedge dx^{i_k} \mapsto 
\cancel{\alpha_k} 
\equiv \frac1{k!} \alpha_{i_1\ldots i_k}\gamma^{i_1\ldots i_k}$ in three dimensions (and, more generally, in any odd dimension). Unlike what happens in even dimensions, the antisymmetrized gamma matrices $\gamma^{i_1\ldots i_k}$ are a redundant basis for bispinors. For example, we see that the slash of the volume form is a number: 
$\cancel{\vol_3}= \sigma^1 \sigma^2 \sigma^3= i$. More generally we have
\begin{equation}\label{eq:sla*}
	\cancel{\alpha} = - i \,\cancel{* \lambda \alpha}.
\end{equation}
In other words, when we identify a form with its image under the Clifford map, we lose some information: we effectively have an equivalence $\alpha \cong -i * \lambda \alpha$. When evaluating $\psi^{1,2}$, we can give the corresponding forms as an even form, or as an odd form, or as a mix of the two. 

Let us first consider $\chi \otimes \chi^\dagger$. We can choose to express it as an odd form. In its Fierz expansion, both its one-form part and its three-form part are a priori non-zero; we can parameterize them as 
\begin{equation}\label{eq:ccd}
	\chi \otimes \chi^\dagger = \frac12 (e_3 - i \vol_3)\ .
\end{equation}
(We can also write this in a mixed even/odd form as $\chi \otimes \chi^\dagger = \frac12(1+e_3)$; recall that the right hand sides have to be understood with a Clifford map applied to them.) $e_3$ is clearly a real vector, whose name has been chosen for later convenience. The fact that the three-form part is simply $-\frac i2 \vol_3$ follows from $|| \chi ||=1$. Notice also that
\begin{equation}\label{eq:e3chi}
	e_3 \chi = \sigma_i \chi e_3^i = \sigma_i \chi \chi^\dagger \sigma^i \chi =  \frac12(-e_3 -3i\vol_3) \chi \quad \Rightarrow \quad e_3 \chi = \chi 
\end{equation}
where we have used (\ref{eq:ccd}), and that $\sigma_i \alpha_k \sigma^i = (-)^k(3-2k) \alpha_k$ on a $k$-form. (\ref{eq:e3chi}) also implies that $e_3$ has norm one.\footnote{An alternative, perhaps more amusing, way of seeing this is to consider $\chi \otimes \chi^\dagger$ as a two-by-two spinorial matrix. It has rank one, which will be true if and only if its determinant is one. Using that $\det(A)=\frac12({\rm Tr}(A)^2-{\rm Tr}(A^2))$ for 2$\times$2 matrices, one gets easily that $e_3$ has norm one.}

Coming now to $\chi\otimes \overline{\chi}$, we notice that the three-form part in its Fierz expansion is zero, since $\overline{\chi}\chi = \chi^t B_3 \chi = 0$. The one-form part is now a priori no longer real; so we write
\begin{equation}\label{eq:ccb}
	\chi \otimes \overline{\chi} = \frac12 (e_1 + i e_2)\ .
\end{equation}
Similar manipulations as in (\ref{eq:e3chi}) show that $(e_1 + i e_2)\chi = 0$; using this, we get that 
\begin{equation}
	e_i \cdot e_j = \delta_{ij}\ .
\end{equation}
In other words, $\{ e_i \}$ is a vielbein, as notation would suggest. 


\subsection{Two spinors} 
\label{sub:two}

We will now analyze the case with two spinors $\chi_1\neq \chi_2$ (again both with norm one). We will proceed in a similar fashion as in \cite[Sec.~3.1]{halmagyi-t}. 

Our aim is to parameterize the bispinors $\psi^{1,2}$ in (\ref{eq:psichi}). 
Let us first consider their zero-form parts, $\chi_2^\dagger \chi_1$ and $\chi_2^{c\,\dagger}\chi_1$. The parameterization (\ref{eq:e3chi}) can be applied to both $\chi_1$ and $\chi_2$, resulting in two one-forms $e_3^i$. (This notation is a bit inconvenient, but these two one-forms will cease to be useful very soon.) Using then (\ref{eq:ccd}) twice, we see that
\begin{equation}
	| \chi_2^\dagger \chi_1 |^2 = \chi_2^\dagger \chi_1 \chi_1^\dagger \chi_2 = {\rm Tr}(\chi_1 \chi_1^\dagger \chi_2 \chi_2^\dagger) = 
	\frac14{\rm Tr}\left((1+e_3^1)(1+e_3^2)\right)= \frac12 (1+ e_3^1 \cdot e_3^2)\ .
\end{equation}
Similarly we have
\begin{equation}
	| \chi_2^{c\,\dagger} \chi_1 |^2 = {\rm Tr}(\chi_1 \chi_1^{c\,\dagger} \chi_2 \chi_2^{c\,\dagger}) = 
	\frac14{\rm Tr}\left((1+e_3^1)(1-e_3^2)\right)= \frac12 (1- e_3^1 \cdot e_3^2)=1-| \chi_2^\dagger \chi_1 |^2\ .
\end{equation}
Both $| \chi_2^\dagger \chi_1 |^2$ and $| \chi_2^{c\,\dagger} \chi_1 |^2$ are positive and $\le 1$. Thus we can parameterize $\chi_2^\dagger \chi_1 = e^{ia} \cos(\psi)$, $\chi_2^{c\,\dagger} \chi_1 = e^{ib} \sin(\psi)$. (The name of this angle should not be confused with the forms $\psi^{1,2}$.) By suitably multiplying $\chi_1$ and $\chi_2$ by two phases, we can assume $a=-\frac{\pi}{2}$ and $b=\frac{\pi}{2}$; we will reinstate generic values of these phases at the very end. Thus we have
\begin{equation}\label{eq:chipsi}
	\chi_2^\dagger \chi_1 = -i\cos(\psi) \ ,\qquad \chi_2^{c\,\dagger} \chi_1 = i\sin(\psi)\ .
\end{equation}

Just as in \cite[Sec.~3.1]{halmagyi-t}, we can now introduce 
\begin{equation}\label{eq:chi0}
	\chi_0 = \frac{1}{2} (\chi_1 - i \chi_2) \ ,\qquad 
	\tilde \chi_0 = \frac{1}{2}(\chi_1 + i \chi_2) \ .
\end{equation}
In three Euclidean dimensions, a spinor and its conjugate form a (pointwise) basis of the space of spinors. For example, $\chi_0$ and $\chi_0^c$ are a basis. We can then expand $\tilde\chi_0$ on this basis. Actually, its projection on $\chi_0$ vanishes, due to (\ref{eq:chipsi}): $\chi_0^\dagger \tilde \chi_0 = \frac i4 (\chi_1^\dagger \chi_2 + \chi_2^\dagger \chi_1)=0$. With a few more steps we get
\begin{equation}\label{eq:tchi0}
	\tilde \chi_0 = \frac{\chi_0^{c\,\dagger} \tilde \chi_0}{|| \chi_0 ||^2} \chi_0^c = \tan\left(\frac{\psi}2\right) \chi_0^c\ .
\end{equation}
We can now invert (\ref{eq:chi0}) for $\chi_1$ and $\chi_2$, and use (\ref{eq:tchi0}). It is actually more symmetric-looking to define $\chi_0 \equiv \cos\left(\frac{\psi}2\right) \chi$, to get
\begin{equation}\label{eq:chi12psi}
	\chi_1 = \cos\left(\frac{\psi}2\right) \chi + \sin\left(\frac{\psi}2\right) \chi^c \ ,\qquad
	\chi_2 = i \left(\cos\left(\frac{\psi}2\right)\chi -\sin\left(\frac{\psi}2\right) \chi^c\right)\ .
\end{equation}
We have thus obtained a parameterization of two spinors $\chi_1$ and $\chi_2$ in terms of a single spinor $\chi$ and of an angle $\psi$. Let us count our parameters, to see if our result makes sense. A spinor $\chi$ of norm 1 accounts for 3 real parameters; $\psi$ is one more. We should also recall we have rotated both $\chi_{1,2}$ by a phase at the beginning of our computation, to make things easier. We have a grand total of 6 real parameters, which is correct for two spinors of norm 1 in three dimensions. 

We can now use the parameterization (\ref{eq:chi12psi}), and the bilinears (\ref{eq:ccd}), (\ref{eq:ccb}) obtained in section \ref{sub:one}:
	\begin{align} \label{eq:chi1chi2pre}
	\chi_1 \otimes \chi_2^\dagger &= -i \left[ \cos^2\left(\frac{\psi}2\right) \chi \chi^\dagger - \sin^2\left(\frac{\psi}2\right) \chi^c \chi^{c\, \dagger} + \cos\left(\frac{\psi}2\right)\sin \left(\frac{\psi}2\right) (\chi^c\chi^\dagger - \chi \chi^{c\,\dagger}) \right] \nonumber \\
	&=-\frac{i}{2} \left[ e_3 - i \sin(\psi) e_2 -i \cos(\psi) \vol_3\right]\ .
	\end{align}
A computation along these lines allows us to evaluate $\chi_1 \otimes \overline{\chi_2} $ as well. We can also reinstate at this point the phases of $\chi_1$ and $\chi_2$, absorbing the overall factor $- i$. The bilinear in (\ref{eq:chi1chi2pre}) is expressed as an odd form, but we will also need its even-form expression; this can be obtained by using (\ref{eq:sla*}). Recalling the definition (\ref{eq:psichi}), we get:
\begin{subequations}\label{eq:psigen}
	\begin{align}
		&\psi^1_+ = \frac{e^{i \theta_1}}2 \left[ \cos(\psi) + e_1\wedge(- i e_2 + \sin(\psi) e_3) \right] \ ,\quad
		\psi^1_- = \frac{e^{i \theta_1}}2 \left[ e_3 - i \sin(\psi) e_2 -i \cos(\psi) \vol_3 \right] \ ; \\
		&\psi^2_+ = \frac{e^{i \theta_2}}2 \left[ \sin(\psi) - (i e_2 + \cos(\psi) e_1)\wedge e_3 \right] \ ,\quad
		\psi^2_- = \frac{e^{i \theta_2}}2 \left[ e_1 + i \cos(\psi) e_2 -i \sin(\psi) \vol_3 \right] \ .
	\end{align}
\end{subequations}
Notice that these satisfy automatically (\ref{eq:73norm}).

Armed with this parameterization, we will now attack the system (\ref{eq:73}) for AdS$_7\times M_3$ solutions. 



\section{General results} 
\label{sec:gen}

In section \ref{sub:73}, we have obtained the system (\ref{eq:73}), equivalent to supersymmetry for AdS$_7\times M_3$ solutions. The $\psi^{1,2}_\pm$ appearing in that system are not just any forms; they should have the property that they can be written as bispinors as in (\ref{eq:psichi}). In section \ref{sub:two}, we have obtained a parameterization for the most general set of $\psi^{1,2}_\pm$ that fulfills that constraint; it is (\ref{eq:psigen}), where $\{e_i\}$ is a vielbein. 

Thus we can now use (\ref{eq:psigen}) into the differential system (\ref{eq:73}), and explore its consequences. 

\subsection{Purely geometrical equations} 
\label{sub:geo}

We will start by looking at the equations in (\ref{eq:73}) that do not involve any fluxes. These are (\ref{eq:73dAR}), and the lowest-component form part of (\ref{eq:73I1}), (\ref{eq:73R1}) and (\ref{eq:732}). 

First of all, we can observe quite quickly that the IIB case cannot possibly work. (\ref{eq:73I1}), (\ref{eq:73R1}) and (\ref{eq:732}) all have a zero-form part coming from their right-hand side, which, using (\ref{eq:psigen}), read respectively 
\begin{equation}\label{eq:iib}
	\cos(\psi) \cos(\theta_1) = 0 \ ,\qquad 
	\cos(\psi) \sin(\theta_1) = 0 \ ,\qquad
	\sin(\psi) e^{i \theta_2}= 0 \ .
\end{equation}  
These cannot be satisfied for any choice of $\psi$, $\theta_1$ and $\theta_2$. Thus we can already exclude the IIB case.\footnote{This quick death is reminiscent of the fate of AdS$_4 \times M_6$ with SU(3) structure in IIB. The system in \cite{gmpt2} has a zero-form equation and two-form equation coming from the right-hand side of its fluxless equation, which look like $\cos(\theta)=0=\sin(\theta)J$, where $\theta$ is an angle similar to $\psi$ in (\ref{eq:psigen}). This is consistent with a no-go found with lengthier computations in \cite{behrndt-cvetic-gao}.}

Having disposed of IIB so quickly, we will devote the rest of the paper to IIA. Actually, we already know that we can get something new only with non-zero Romans mass, $F_0\neq 0$. This is because for $F_0=0$ we can lift to an eleven-dimensional supergravity solution AdS$_7 \times N_4$. There, we only have a four-form flux $G_4$ at our disposal, and the only way not to break the SO(6,2) invariance of AdS$_7$ is to switch it on along the internal four-manifold $N_4$. This is the Freund--Rubin Ansatz, which requires $N_4$ to admit a Killing spinor. This means that the cone $C(N_4)$ over $N_4$ admits a covariantly constant spinor; but in five dimensions the only manifold with restricted holonomy is $\rr^5$ (or one of its orbifolds, of the form $\rr^4/\Gamma \times \rr$). Thus we know already that all solutions with $F_0=0$ lift to AdS$_7\times S^4$ (or AdS$_7\times S^4/\Gamma$) in eleven dimensions. (In fact we will see later how AdS$_7\times S^4$ reduces to ten dimensions.) We will thus focus on $F_0\neq0$, and use the case $F_0=0$ as a control. 

In IIA, the lowest-degree equations of (\ref{eq:73I1}), (\ref{eq:73R1}) and (\ref{eq:732}) are one-forms; they are less dramatic than (\ref{eq:iib}), but still rather interesting. Using (\ref{eq:psigen}), after some manipulations we get
\begin{equation}\label{eq:ei}
\begin{split}
	&e_1 = -\frac14 e^A \sin(\psi) d \theta_2 \ ,\qquad 
	e_2 = \frac14 e^A (d \psi + \tan(\psi)d(5A-\phi) ) \ ,\\
	&e_3 = \frac14 e^A \left(-\cos(\psi) d \theta_1 + \frac{\cot(\theta_1)}{\cos(\psi)}d(5A-\phi)\right) \ ,
\end{split}
\end{equation}
and
\begin{equation}\label{eq:psisol}
	x dx = (1+x^2) d \phi - (5+x^2) dA \ ,
\end{equation}
where 
\begin{equation}\label{eq:x}
	x \equiv \cos(\psi) \sin(\theta_1)\ ,
\end{equation}
and we have dropped the subscript $_3$ on the warping function: $A\equiv A_3$ from now on.
Notice that (\ref{eq:ei}) determine the vielbein. Usually (i.e.~in other dimensions), the geometrical part of the differential system coming from supersymmetry gives the \emph{derivative} of the forms defining the metric. In this case, the forms themselves are determined in terms of derivatives of the angles appearing in our parameterizations. This will allow us to give a more complete and concrete classification than is usually possible. 

We still have (\ref{eq:73dAR}). Notice that (\ref{eq:73I1}) allows to write it as $dA \wedge d(e^{3A-\phi}x)=0$. Using also (\ref{eq:psisol}), we get 
\begin{equation}
	dA \wedge d \phi = 0 \ .
\end{equation} 
This means that $\phi$ is functionally dependent on $A$:\footnote{(\ref{eq:phiA}) excludes the case where $A$ is constant in a region. However, it is easy to see that this case cannot work. Indeed, in this case (\ref{eq:psisol}) can be integrated as $e^\phi \propto\sqrt{1-x^2}$, which is incompatible with (\ref{eq:f0}) below.}
\begin{equation}\label{eq:phiA}
	\phi = \phi(A)\ .
\end{equation}
(\ref{eq:psisol}) then means that $x$ too is functionally dependent on $A$: $x=x(A)$. 


\subsection{Fluxes} 
\label{sub:flux}

So far, we have analyzed (\ref{eq:73dAR}), and the one-form part of (\ref{eq:73I1}), (\ref{eq:73R1}) and (\ref{eq:732}). Before we look at their three-form part too, it is convenient to look at (\ref{eq:73f}), which gives us the RR flux, for reasons that will become apparent. 

First we compute $F_0$ from (\ref{eq:73f}): 
\begin{equation}\label{eq:f0}
	F_0 = 4 x e^{-A-\phi} \frac{3-\del_A \phi}{5-2x^2 - \del_A \phi}\ . 
\end{equation}
The Bianchi identity for $F_0$ says that it should be (piecewise) constant. It will thus be convenient to use (\ref{eq:f0}) to eliminate $\del_A \phi$ from our equations.

Before we go on to analyze our equations, let us also introduce the new angle $\beta$ by
\begin{equation}\label{eq:beta}
	\sin^2(\beta) = \frac{\sin^2(\psi)}{1-x^2}\ . 
\end{equation}
We can now use $x$ as defined in (\ref{eq:x}) to eliminate $\theta_1$, and $\beta$ to eliminate $\psi$. This turns out to be very convenient in the following, especially in our analysis of the metric in section \ref{sub:metric} below (which was our original motivation to introduce $\beta$). 

After these preliminaries, let us give the expression for $F_2$ as one obtains it from (\ref{eq:73f}):
\begin{equation}\label{eq:f2}
	F_2=\frac1{16}\sqrt{1-x^2}e^{A-\phi}(x e^{A+\phi} F_0  - 4) \vol_{S^2} \ ,
\end{equation}
where
\begin{equation}\label{eq:vol2}
	\vol_{S^2}= \sin(\beta) d \beta \wedge d\theta_2
\end{equation}
is formally identical to the volume form for a round $S^2$ with coordinates $\{ \beta, \theta_2 \}$. We will see later that this is no coincidence.

Finally, let us look at the three-form part of (\ref{eq:73I1}), (\ref{eq:73R1}) and (\ref{eq:732}). One of them can be used to determine $H$:
\begin{equation}\label{eq:H}
	H= \frac18 e^{2A}\sqrt{1-x^2} \,\frac{6+x F_0 e^{A+\phi}}{4+x F_0 e^{A+\phi}}\, dx \wedge \vol_{S^2} \ , 
\end{equation}
while the other two turn out to be identically satisfied. 

Our analysis is not over: we should of course now impose the equation of motion, and the Bianchi identities for our fluxes. The equation of motion for $F_2$, $d*F_2 + H *F_0 = 0$, follows automatically from (\ref{eq:73f}), much as it happens in the pure spinor system for AdS$_4 \times M_6$ solutions \cite{gmpt2}. We should then impose the Bianchi identity for $F_2$, which reads $d F_2 - H F_0=0$ (away from sources). This does not follow manifestly from (\ref{eq:73f}), but in fact it is a consequence of the explicit expressions (\ref{eq:f0}), (\ref{eq:f2}) and (\ref{eq:H}) above. When $F_0\neq 0$, it also implies that the $B$ field such that $H=dB$ can be locally written as 
\begin{equation}\label{eq:Bb}
	B_2 = \frac{F_2}{F_0}+ b
\end{equation}
for a closed two-form $b$. Using a gauge transformation, it can be assumed to be proportional (by a constant) to $\vol_{S^2}$; we then have that it is a constant, $\del_A b=0$. 

The equation of motion for $H$, which reads for us $d(e^{7A-2 \phi} *_3 H)= e^{7A} F_0 *_3 F_2$ (again away from sources), is also automatically satisfied, as shown in general in \cite{koerber-tsimpis}. Finally, since we have checked all the conditions for preserved supersymmetry, the Bianchi identities and the equations of motion for the fluxes, the equations of motion for the dilaton and for the metric will now follow \cite{lust-tsimpis}.


\subsection{The system of ODEs} 
\label{sub:ode}

Let us now sum up the results of our analysis of (\ref{eq:73}). Most of our equations determine some fields: (\ref{eq:ei}) give the vielbein, and (\ref{eq:f0}), (\ref{eq:f2}), (\ref{eq:H}) give the fluxes. The only genuine differential equations we have are (\ref{eq:psisol}), and the condition that $F_0$ should be constant. Recalling that $\phi$ is functionally dependent on $A$, (\ref{eq:phiA}), these two equations can be written as
\begin{subequations}\label{eq:odeA}
\begin{align}
	&\del_A \phi = 5 - 2 x^2 + \frac{8 x(x^2-1)}{4 x - F_0 e^{A+\phi}}\ , 
	\\
	&\del_A x = 2(x^2-1)\,\frac{x e^{A+\phi} F_0 + 4}{4x - F_0 e^{A+\phi}}\ .
\end{align}	
\end{subequations}

We thus have reduced the existence of a supersymmetric solution of the form AdS$_7 \times M_3$ in IIA to solving the system of ODEs (\ref{eq:odeA}). It might look slightly unsettling that we are essentially using at this point $A$ as a coordinate, which might not always be a wise choice (since $A$ might not be monotonic). For that matter, our analysis has so far been completely local; we will start looking at global issues in section \ref{sub:metric}, and especially \ref{sub:top}.

\begin{figure}[ht]
	\centering
		\includegraphics[scale=1]{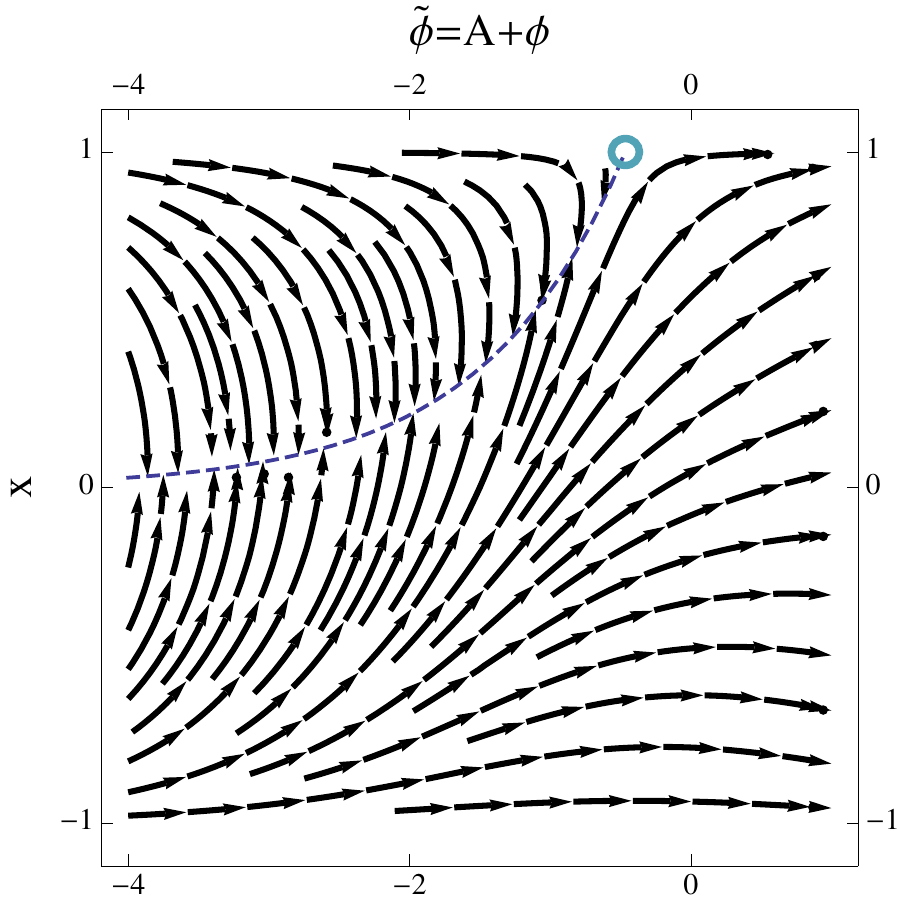}
	\caption{A plot of the vector field induced by (\ref{eq:odeA}) on $\{\tilde \phi \equiv \phi + A, x\}$, for $F_0=40/2\pi$ (in agreement with flux quantization, (\ref{eq:F0n0}) below). The green circle represents the point $\{\phi+A=\log(4/F_0),x=1\}$, whose role will become apparent in section \ref{sub:loc}. The dashed line represents the locus along which the denominators in (\ref{eq:odeA}) vanish.}
	\label{fig:stream}
\end{figure}

Unfortunately we have not been able to find analytic solutions to (\ref{eq:odeA}), other than in the $F_0=0$ case (which we will see in section \ref{sub:massless}). For the more interesting $F_0 \neq 0$ case, we can gain some intuition by noticing that the system becomes autonomous (i.e.~it no longer has explicit dependence on the ``time'' variable $A$) if one defines $\tilde \phi \equiv \phi + A$. The system for $\{ \del_A \tilde \phi, \del_A x \}$ can now be thought of as a vector field in two dimensions; we plot it in figure \ref{fig:stream}.

We will study the system (\ref{eq:odeA}) numerically in section \ref{sec:exp}. Before we do that, we should understand what boundary conditions we should impose. We will achieve this by analyzing global issues about our setup, that we have so far ignored.


\subsection{Metric} 
\label{sub:metric}

The metric
\begin{equation}\label{eq:eaea}
	ds^2_{M_3} = e_a e_a
\end{equation}
following from (\ref{eq:ei}) looks quite complicated. However, it simplifies enormously if we rewrite it in terms of $\beta$ in (\ref{eq:beta}):\footnote{In fact, the definition of $\beta$ was originally found by trying to understand the global properties of the metric (\ref{eq:eaea}). Looking at a slice $x=$const, one finds that the metric in $\{\theta_1,\theta_2\}$ has constant positive curvature; the definition of $\beta$ becomes then natural. Nontrivially, this definition also gets rid of non-diagonal terms of the type $dA d \theta_1$ that would arise from (\ref{eq:ei}).}
\begin{equation}\label{eq:metricfibr}
	ds^2_{M_3}=e^{2A}(1-x^2)\left[\frac{16}{(4x-e^{A+\phi}F_0)^2}\,dA^2+ \frac1{16}\,ds^2_{S^2} \right] \ ,\qquad
	ds^2_{S^2}= d \beta^2 + \sin^2(\beta) d\theta_2^2\ .
\end{equation}
\emph{Without any Ansatz}, the metric has taken the form of a fibration of a round $S^2$, with coordinates $\{\beta, \theta_2\}$, over an interval with coordinate $A$. Notice that none of the scalars appearing in (\ref{eq:metricfibr}) (and indeed in the fluxes (\ref{eq:f0}), (\ref{eq:f2}), (\ref{eq:H})) were originally intended as coordinates, but rather as functions in the parameterization of the pure spinors $\psi^{1,2}$. Usually, one would then need to introduce coordinates independently, and to make an Ansatz about how all functions should depend on those coordinates, sometimes imposing the presence of some particular isometry group in the process. 

Here, on the other hand, the functions we have introduced are suggesting themselves as coordinates to us rather automatically. Since so far our expressions for the metric and fluxes were local, we are free to take their suggestion. We will take $\beta$ to be in the range $[0,\pi]$, and $\theta_2$ to be periodic with period $2 \pi$, so that together they describe an $S^2$ as suggested by (\ref{eq:metricfibr}), and also by the two-form (\ref{eq:vol2}) that appeared in (\ref{eq:f2}), (\ref{eq:H}).\footnote{A slight variation is to take $\rr\pp^2=S^2/\zz_2$ instead of $S^2$; this will not play much of a role in what follows, except for some solutions with O6-planes that we will mention in sections \ref{sub:massless} and \ref{sub:nod8}.}

It is not hard to understand why this $S^2$ has emerged. The holographic dual of any solutions we might find is a $(1,0)$ CFT in six dimensions. Such a theory would have SU(2) R-symmetry; an SU(2) isometry group should then appear naturally on the gravity side as well. This is what we are seeing in (\ref{eq:metricfibr}). 

The fact that the $S^2$ in (\ref{eq:metricfibr}) is rotated by R-symmetry also helps to explain a possible puzzle about IIB. Often, given a IIA solution, one can produce a IIB one via T-duality along an isometry. All the Killing vectors of the $S^2$ in (\ref{eq:metricfibr}) vanish in two points; T-dualizing along any such direction would produce a non-compact solution in IIB, but still a valid one. But the IIB case died very quickly in section \ref{sub:geo}; there are no solutions, not even non-compact or singular ones. Here is how this puzzle is resolved. Since the SU(2) isometry group of the $S^2$ is an R-symmetry, supercharges transform as a doublet under it (we will see this more explicitly in section \ref{sub:su2}). Thus even the strange IIB geometry produced by T-duality along a U(1) isometry of $S^2$ would not be supersymmetric. 

Even though we have promoted $\beta$ and $\theta_2$ to coordinates, it is hard to do the same for $A$, which actually enters in the seven-dimensional metric (see (\ref{eq:met73})). We would like to be able to cover cases where $A$ is non-monotonic. One possibility would be to use $A$ as a coordinate piecewise. We find it clearer, however, to introduce a coordinate $r$ defined by $dr = 4e^A \frac{\sqrt{1-x^2}}{4x-e^{A+\phi}F_0}dA$, so that the metric now reads
\begin{equation}\label{eq:met-r}
	ds^2_{M_3} = dr^2 + \frac1{16}e^{2A}(1-x^2)ds^2_{S^2} \ .
\end{equation}
In other words, $r$ measures the distance along the base of the $S^2$ fibration. Now $A$, $x$ and $\phi$ have become functions of $r$. From  (\ref{eq:odeA}) and the definition of $r$ we have 
\begin{equation}\label{eq:oder}
\begin{split}
	&\del_r \phi = \frac14 \frac{e^{-A}}{\sqrt{1-x^2}} (12 x + (2x^2-5)F_0 e^{A+\phi}) \ ,\\
	&\del_r x = -\frac12 e^{-A}\sqrt{1-x^2} (4+x F_0 e^{A+\phi}) \ ,\\
	&\del_r A = \frac14 \frac{e^{-A}}{\sqrt{1-x^2}} (4x - F_0 e^{A+\phi})\ .
\end{split}
\end{equation}
We have introduced a square root in the system, but notice that $-1\le x\le 1$ already follows from requiring that $ds^2_{M_3}$ in (\ref{eq:metricfibr}) has positive signature. (We choose the positive branch of the square root.)

Let us also record here that the NS three-form also simplifies in the coordinates introduced in this section: 
\begin{equation}\label{eq:Hnice}
	H = - (6 e^{-A} + x F_0 e^\phi) {\rm vol}_3\ ,
\end{equation}
where ${\rm vol}_3$ is the volume form of the metric $ds^2_{M_3}$ in (\ref{eq:met-r}) or (\ref{eq:metricfibr}).

We have obtained so far that the metric is the fibration of an $S^2$ (with coordinates $(\beta, \theta_2)$) over a one-dimensional space. The SU(2) isometry group of the $S^2$ is to be identified holographically with the R-symmetry group of the $(1,0)$-superconformal dual theory.
For holographic applications, we would actually like to know whether the total space of the $S^2$-fibration can be made compact. We will look at this issue in section \ref{sub:top}. Right now, however, we would like to take a small detour and see a little more clearly how the R-symmetry SU(2) emerges in the pure spinors $\psi^{1,2}$.


\subsection{${\rm SU}(2)$-covariance} 
\label{sub:su2}

We have just seen that the metric takes the particularly simple form (\ref{eq:met-r}) in coordinates $(r,\beta,\theta_2)$; the appearance of the $S^2$ is related to the SU(2) R-symmetry group of the $(1,0)$ holographic dual.

Since these coordinates are so successful with the metric, let us see whether they also simplify the pure spinors $\psi^{1,2}$. We can start by the zero-form parts of (\ref{eq:psigen}), which read
\begin{equation}\label{eq:psi0}
	\psi^1_0= i x + \sqrt{1-x^2}\, \cos(\beta) \ ,\qquad \psi^2_0 = \sqrt{1-x^2}\, \sin(\beta) e^{i \theta_2}\ .
\end{equation}
Recalling that $(\beta, \theta_2)$ are the polar coordinates on the $S^2$ (see the expression of $ds^2_{S^2}$ in (\ref{eq:metricfibr})), we recognize in (\ref{eq:psi0}) the appearance of the $\ell=1$ spherical harmonics 
\begin{equation}
	y^\alpha = \{ \sin(\beta) \cos(\theta_2), \sin(\beta) \sin(\theta_2), \cos(\beta) \}\ .
\end{equation}
Notice that $y^3$ appears in $\psi^1= \chi_1 \otimes \chi_2^\dagger$, while $y^1+i y^2$ appears in $\psi^2=\chi_1 \otimes \chi_2^{c\,\dagger}$. This suggests that we introduce a 2$\times$2 matrix of bispinors. From (\ref{eq:eps73}) we see that for IIA ${\chi_1 \choose \chi_1^c}$ and ${\chi_2 \choose -\chi_2^c}$ are both SU(2) doublets, so that it is natural to define 
\begin{equation}
	\Psi= \left( \begin{array}{c}
		\chi_1 \\ \chi_1^c	\end{array}\right) \otimes 
		(\chi_2^\dagger \ , -\chi_2^{c\,\dagger}) =  
		\left( \begin{array}{cc}
			\psi^1  & \psi^2\\
			(-)^{\rm deg}(\psi^2)^* & -(-)^{\rm deg}(\psi^1)^*
		\end{array}\right)\ ,
\end{equation}
where $(-)^{\rm deg}$ acts as $\pm$ on a even (odd) form. The even-form part can then be written as
\begin{subequations}\label{eq:PSI}
\begin{equation}\label{eq:PSI0}
	\Psi^{ab}_+ = i {\rm Im} \psi^1_+ \, {\rm Id}_2 + \left( {\rm Re}  \psi^2_+ \sigma_1 - {\rm Im}  \psi^2_+  \sigma_2 + {\rm Re}  \psi^1_+ \sigma_3 \right)\ ,
\end{equation}
where $\sigma_\alpha$ are the Pauli matrices while the odd-form part is 
\begin{equation}\label{eq:PSI1}
	\Psi^{ab}_- = {\rm Re} \psi^1_- \, {\rm Id}_2 + i\left( {\rm Im}  \psi^2_- \sigma_1 + {\rm Re}  \psi^2_- \sigma_2 + {\rm Im}  \psi^1_- \sigma_3\right) \ .
\end{equation}
\end{subequations}
(\ref{eq:PSI}) shows more explicitly how the R-symmetry SU(2) acts on the bispinors $\Psi^{ab}$, which split between a singlet and a triplet. If we go back to our original system (\ref{eq:73}), we see that (\ref{eq:73I1}), (\ref{eq:73f}), (\ref{eq:73dAR}) each behave as a singlet, while (\ref{eq:73R1}), (\ref{eq:732}) behave as a triplet --- thanks also to the fact that the factor $e^{5A - \phi}$ appears in both those equations. 

More concretely, (\ref{eq:psi0}) can now be written as
\begin{subequations}\label{eq:Psi}
\begin{equation}\label{eq:Psi0}
	\Psi^{ab}_0= i x\, {\rm Id}_2 + \sqrt{1-x^2}\, y^\alpha \sigma_\alpha\ ;
\end{equation}
the one-form part reads
\begin{equation}\label{eq:Psi1}
	\Psi^{ab}_1 = \sqrt{1-x^2} dr\, {\rm Id}_2 + i \left[ x y^\alpha dr + \frac14 e^A \sqrt{1-x^2} \, dy^\alpha \right] \sigma_\alpha   \ .
\end{equation}
\end{subequations}
The rest of $\Psi^{ab}$ can be determined by (\ref{eq:sla*}): $\Psi^{ab}_3=-i*_3 \Psi^{ab}_0= -i \Psi^{ab}_0 \,\vol_3$, $\Psi^{ab}_2=-i*_3 \Psi^{ab}_1$. (The three-dimensional Hodge star can be easily computed from (\ref{eq:met-r}).)

We will now turn to the global analysis of the metric (\ref{eq:met-r}). 


\subsection{Topology} 
\label{sub:top}

We now wonder whether the $S^2$ fibration in (\ref{eq:metricfibr}) can be made compact.

One possible strategy would be for $r$ to be periodically identified, so that the topology of $M_3$ would become $S^1\times S^2$. This is actually impossible: from (\ref{eq:oder}) we have
\begin{equation}\label{eq:noninc}
	\del_r(x e^{3A-\phi}) = -2 \sqrt{1-x^2} e^{2A-\phi} \le 0 \ .
\end{equation}
This can also be derived quickly from (\ref{eq:73I1}) using the singlet part of (\ref{eq:Psi}). Now, $x e^{3A-\phi}$ is continuous;\footnote{This might not be fully obvious in presence of D8-branes, but we will see later that it is true even in that case, basically because $\phi$ is a physical field, and $A$ and $x$ appear as coefficients in the metric.} for $r$ to be periodically identified, $x e^{3A-\phi}$ should be a periodic function. However, thanks to (\ref{eq:noninc}), it is nowhere-increasing. It also cannot be constant, since $x$ would be $\pm 1$ for all $r$, which makes the metric in (\ref{eq:metricfibr}) vanish. Thus $r$ cannot be periodically identified. 

We then have to look for another way to make $M_3$ compact. The only other possibility is in fact to shrink the $S^2$ at two values of $r$, which we will call $r_{\rm N}$ and $r_{\rm S}$; the topology of $M_3$ would then be $S^3$. The subscripts stand for ``north'' and ``south''; we can visualize these two points as the two poles of the $S^3$, and the other, non-shrunk copies of $S^2$ over any $r\in (r_{\rm N}, r_{\rm S})$ to be the ``parallels'' of the $S^3$. Of course, since (\ref{eq:oder}) does not depend on $r$, we can assume without any loss of generality that $r_{\rm N}=0$.

We will now analyze this latter possibility in detail.


\subsection{Local analysis around poles} 
\label{sub:loc}

We have just suggested to make $M_3$ compact by having the $S^2$ fiber over an interval $[r_{\rm N},r_{\rm S}]$, and by shrinking it at the two extrema. In this case $M_3$ would be homeomorphic to $S^3$.

To realize this idea, from (\ref{eq:met-r}) we see that $x$ should go to $1$ or $-1$ at the two poles $r_{\rm N}$ and $r_{\rm S}$. To make up for the vanishing of the $\sqrt{1-x^2}$'s in the denominators in (\ref{eq:oder}), we should also make the numerators vanish. This is accomplished by having $e^{A+\phi}= \pm 4/F_0$ at those two poles (which is obviously only possible when $F_0 \neq 0$). We can now also see that $\del_r x \sim -4 e^{-A}\sqrt{1-x^2}\le 0 $ around the poles. Since, as we noticed earlier, $-1\le x \le 1$, $x$ should actually be 1 at $r_{\rm N}$, and $-1$ at $r_{\rm S}$. Summing up:
\begin{equation}\label{eq:bc}
	\left\{x=1,\ e^{A+\phi}=\frac4{F_0}\right\} \ \ {\rm at}\ r=r_{\rm N} \ ,\qquad
	\left\{x=-1,\ e^{A+\phi}=-\frac4{F_0}\right\} \ \ {\rm at}\ r=r_{\rm S} \ .
\end{equation}

Since we made both numerators and denominators in (\ref{eq:oder}) vanish at the poles, we should be careful about what happens in the vicinity of those points. We want to study the system around the boundary conditions (\ref{eq:bc}) in a power-series approach. (The same could also be done directly with (\ref{eq:odeA}).) Let us first expand around $r_{\rm N}$. As mentioned earlier, thanks to translational invariance in $r$ we can assume $r_{\rm N}=0$ without any loss of generality. We get
\begin{equation}\label{eq:expN}
\begin{split}
	&\phi= -A_0^+ + \log\left(\frac4{F_0}\right)-5 e^{-2A_0^+} r^2 +\frac{172}9 e^{-4 A_0^+}r^4 +O(r)^6 \ ,\\
	&x= 1-8 e^{-2A_0^+} r^2 +\frac{400}9 e^{-4 A_0^+}r^4+O(r)^6\ ,\\
	&A= A_0^+ -\frac13 e^{-2A_0^+} r^2 -\frac{4}{27} e^{-4 A_0^+}r^4+O(r)^6\ .
\end{split}
\end{equation}
$A_0^+$ here is a free parameter. The way it appears in (\ref{eq:expN}) is explained by noticing that (\ref{eq:oder}) is symmetric under
\begin{equation}
	A \to A + \Delta A \ ,\qquad \phi \to \phi - \Delta A \ ,\qquad x \to x \ ,\qquad 
	r \to e^{\Delta A} r \ .
\end{equation}

Applying (\ref{eq:expN}) to (\ref{eq:met-r}), and setting for a moment $r_{\rm N}=0$, we find that the metric has the leading behavior
\begin{equation}\label{eq:metpole}
	ds^2_{M_3} = dr^2 + r^2 ds^2_{S^2} + O(r)^4 = ds^2_{\rr^3} + O(r)^4\ .
\end{equation}
This means that the metric is regular around $r=r_{\rm N}$. The expansion of the fluxes (\ref{eq:f2}), (\ref{eq:H}) is
\begin{equation}\label{eq:fluxpole}
	F_2 = -\frac{10}3 F_0 e^{-A_0^+} r^3 \vol_{S^2} +  O(r)^5 \ ,\qquad 
	H= -10 e^{-A_0^+} r^2 dr \wedge \vol_{S^2} + O(r)^3 \ .
\end{equation}
As for the $B$ field, recall that it can be written as in (\ref{eq:Bb}). (\ref{eq:fluxpole}) shows that around $r=r_{\rm N}=0$, the term $F_2/F_0$ is regular as it is, without the addition of $b$; this suggests that one should set $b=0$. To make this more precise, consider the limit
\begin{equation}
	\lim_{r\to 0} \int_{\Delta_r} H = \lim_{r\to 0} \int_{S^2_r} B_2\ 
\end{equation}
where $\Delta_r$ is a three-dimensional ball such that $\del \Delta_r = S^2_r$. In (\ref{eq:Bb}), the first term goes to zero because $x\to 1$; so the limit is equal to $\int_{S^2} b$, which is constant. This constant signals a delta in $H$. So we are forced to conclude that
\begin{equation}\label{eq:b0}
	b=0 
\end{equation}
near the pole. (However, we will see in section \ref{sub:d8} that $b$ can become non-zero if one crosses a D8 while going away from the pole.)  

To be more precise, (\ref{eq:b0}) should be understood up to gauge transformations. $B$ is not a two-form, but a `connection on a gerbe', in the sense that it transforms non-trivially on chart intersections: on $U\cap U'$, $B_U - B_{U'}$ can be a `small' gauge transformation $d \lambda$, for $\lambda$ a 1-form, or more generally a `large' gauge transformation, namely a two-form whose periods are integer multiples of $4\pi^2$. In our case, if we cover $S^3$ with two patches $U_{\rm N}$ and $U_{\rm S}$, around the equator we can have $B_{\rm N}- B_{\rm S}= N\pi {\rm vol}_{S^2}$. In this case $\int_{S^3} H = B_{\rm N}- B_{\rm S}= N\pi {\rm vol}_{S^2} = (4\pi^2) N$, in agreement with flux quantization for $H$. Thus $b=0$ is also gauge equivalent to any integer multiple of $\pi {\rm vol}_{S^2}$. In practice, however, we will prefer to work with $b=0$ around the poles, and perform a gauge transformation whenever
\begin{equation} \label{eq:hatb}
	\hat b (r) \equiv \frac1{4\pi} \int_{S^2_r} B_2 
\end{equation}
gets outside the ``fundamental region'' $[0,\pi]$. In other words, we will consider $\hat b$ to be a variable with values in $[0,\pi]$, and let it begin and end at 0 at the two poles. $\hat b$ will then wind an integer number $N$ of times around $[0,\pi]$, and this will make sure that $\int_{S^3} H = (4\pi^2)N$, thus taking care of flux quantization for $H$.

So far we have discussed the expansion around the north pole; a similar discussion holds for the expansion around the south pole $r_{\rm S}$. The expressions that replace (\ref{eq:expN}), (\ref{eq:metpole}), (\ref{eq:fluxpole}) can be obtained by using the symmetry of (\ref{eq:oder}) under
\begin{equation}\label{eq:F0-F0}
	x\to - x \ ,\qquad F_0 \to - F_0 \ ,\qquad r \to -r \ .
\end{equation}
The free parameter $A_0^+$ can now be changed to a possibly different free parameter $A_0^-$. 

We have hence checked that the boundary conditions (\ref{eq:bc}) are compatible with our system (\ref{eq:oder}), and that they give rise to a regular metric at the poles. 


\subsection{D8} 
\label{sub:d8}

There is one more ingredient that we will need in section \ref{sec:exp} to exhibit compact solutions: brane sources. In presence of branes the metric cannot be called regular: their gravitational backreaction will give rise to a singularity. A random singularity would call into question the validity of a solution, since the curvature and possibly the dilaton\footnote{In presence of Romans mass, the string coupling is bounded by the inverse radius of curvature in string units: $e^\phi \lsim \frac{l_s}{R_{\rm curv}}$, and is actually generically of the order of the bound \cite{ajtz}.} would diverge there, making the supergravity approximation untrustworthy. We are however sure of the existence of D-branes, in spite of the singularities in their geometry, because we have an open string realization for them. 

D8-branes in particular are even more benign, in a way, because the singularity manifests itself simply as a discontinuity in the derivatives of the coefficients in the metric. In general relativity, such a discontinuity would be subject to the so-called Israel junction conditions \cite{israel-junction}, which are a consequence of the Einstein equations. As we mentioned earlier, in our case, however, supersymmetry guarantees that the equations of motion for the dilaton and metric are automatically satisfied \cite{lust-tsimpis}. Hence, the conditions on the first derivatives will follow from imposing continuity of the fields and supersymmetry. 

Let us be more concrete. We will suppose we have a stack of $n_{\mathrm{D}8}$ D8-branes, possibly with a worldvolume gauge field-strength $f_2$ (not to be confused with the RR field-strength $F_2$), which induces a D6-brane charge distribution on it. The Bianchi identity for such an object reads
\begin{equation}\label{eq:bianchi}
	d_H F = \frac1{2\pi} n_{\mathrm{D}8}e^{\cal F} \delta 
	\qquad \Rightarrow \qquad d \tilde F = \frac1{2\pi} n_{\mathrm{D}8} e^{2\pi f_2} \delta \qquad (\delta\equiv dr \delta(r))\ .
\end{equation}
As usual ${\cal F}= B_2+ 2\pi f_2$; recall from section \ref{sub:73} that $F=F_0+F_2$; and likewise we have defined
\begin{equation}
	\tilde F \equiv e^{-B_2} F = F_0 + (F_2 - B_2 F_0)\ .
\end{equation}
In other words, $\tilde F = F_0 + \tilde F_2$, with $\tilde F_2 = F_2 - B_2 F_0$.
Since $\tilde F_2$ is closed away from sources, it makes sense to define
\begin{equation}\label{eq:n2}
	n_2 \equiv \frac1{2\pi} \int_{S^2} \tilde F_2\ .
\end{equation}
Flux quantization then requires $n_2$ to be an integer, and that
\begin{equation}\label{eq:F0n0}
	F_0 = \frac{n_0}{2\pi}\ ,
\end{equation}
with $n_0$ an integer. (We are working in string units where $l_s=1$.) Integrating now (\ref{eq:bianchi}) across the magnetized stack of D8's gives
\begin{equation}\label{eq:DeltaF2}
	\Delta n_0 = n_{\mathrm{D}8}\ ,\qquad \Delta \tilde F_2 = f_2 \Delta n_0 \ .
\end{equation}

All physical fields should be continuous across the D8 stack. For example, $\Delta \phi=0$. Also, the coefficients of the metric should not jump; in particular, from (\ref{eq:met73}), we see that $\Delta A=0$. Also, since $x$ appears in front of $ds^2_{S^2}$ in (\ref{eq:met-r}), we should have $\Delta x =0$.

Imposing that the $B$ field does not jump is trickier. A first caveat is that $B$ would actually be allowed to jump by a gauge transformation, as discussed in section \ref{sub:loc}. However, we find it less confusing to put the intersection between the charts $U_{\rm N}$ and $U_{\rm S}$ away from the D8's, and to treat $\int_{S^2} B_2$ as a periodic variable as described in section \ref{sub:loc}. 

Thus we will simply impose that $B$ does not jump. First, recall that it can be written as in (\ref{eq:Bb}), when $F_0 \neq 0$. The $b$ term was shown in (\ref{eq:b0}) to be vanishing near the pole, but we will soon see that this conclusion is not valid between D8's. In fact, it is connected to the flux integer $n_2$ defined in (\ref{eq:n2}): from (\ref{eq:Bb}) we have
\begin{equation}
	\tilde F_2 = - F_0 b \ ;
\end{equation}
integrating this on $S^2$, we get $2\pi n_2 = - F_0 \int_{S^2} b$, or in other words
\begin{equation}\label{eq:bn2}
	b=-\frac{n_2}{2 F_0} \vol_{S^2}\ . 
\end{equation}
We can use our result (\ref{eq:f2}) for $F_2$; for this section, it will be convenient to define 
\begin{equation}\label{eq:pq}
	p\equiv\frac1{16}x\sqrt{1-x^2}e^{2A} \ ,\qquad q \equiv \frac14 \sqrt{1-x^2}e^{A-\phi}\ ,
\end{equation}
so that
\begin{equation}\label{eq:f2pq}
	F_2 = (p F_0 - q) \vol_{S^2} \ .
\end{equation}
From this and (\ref{eq:bn2}) we now have
\begin{equation}\label{eq:Bmassive}
	B_2 = \left(p - \frac q{F_0} -\frac{n_2}{2 F_0}\right) \vol_{S^2}\ .
\end{equation}
Let us call $n_0$, $n_2$ the flux integers on one side of the D8 stack, and $n_0'$, $n_2'$ the fluxes on the other side. Let us at first assume that both $n_0$ and $n_0'$ are non-zero. Then, equating $B$ on the two sides, we see that $p$ cancels out, and we get
\begin{equation}\label{eq:q/n0}
	\frac1{n_0}\left( q+ \frac12 n_2\right)= \frac1{n_0'}\left( q+ \frac12 n_2'\right)\ ,
\end{equation}
or in other words
\begin{equation}\label{eq:jump}
	 q|_{r=r_{\mathrm{D}8}} = \frac{n_2' n_0 - n_2 n_0'}{2 (n_0'-n_0)}\ ,
\end{equation}
with $q$ as defined in (\ref{eq:pq}). Notice that, in (\ref{eq:Bb}), the term $F_2/F_0$ and $b$ can both separately jump, while the whole $B_2$ is staying continuous. For this reason, as we anticipated in section \ref{sub:loc}, the conclusion $b=0$ (which implies $n_2=0$ by (\ref{eq:bn2})) will hold near the poles, but can cease to hold after one crosses a D8. (\ref{eq:jump}) is also satisfying in that it is symmetric under exchange $\{n_0,n_2\}  \leftrightarrow \{ n_0',n_2'\}$. Notice also that, under a gauge transformation for the $B$ field, $n_2\to n_2 + n_0 \Delta B $, $n_2' \to n_2' + n_0' \Delta B$, and (\ref{eq:jump}) remains unchanged.

A constraint on the discontinuity should also come from the $F_2$ Bianchi identity (\ref{eq:bianchi}). Using (\ref{eq:f2pq}), we see that the only discontinuities are coming from the jump in $F_0$, so that we get
\begin{equation}
	d_H F = \Delta F_0 (1 + p \vol_{S^2}) \delta = \Delta F_0 \,e^{p \vol_{S^2}} \delta \ .
\end{equation}
Comparing this with (\ref{eq:bianchi}) we see that ${\cal F}= p \vol_{S^2}$. It also follows that
\begin{equation}\label{eq:dtF2}
	d \tilde F_2 = \Delta F_0 (-B_2 + p \vol_{S^2}) \delta = 
	\frac{\Delta F_0}{F_0} \left(q + \frac12 n_2\right) \vol_{S^2}\ .
\end{equation}
The expression on the right-hand side is not ambiguous thanks to (\ref{eq:f2pq}). Comparing (\ref{eq:dtF2}) with (\ref{eq:bianchi}) again, we see that $f_2=\frac1{F_0}\left(q+\frac{n_2}2\right)$. Going back to (\ref{eq:DeltaF2}), we learn that
\begin{equation}
	\frac{\Delta n_2}{\Delta n_0 } = \frac1{n_0}\left(q+ \frac12 n_2\right)\ .
\end{equation}
This is actually nothing but (\ref{eq:jump}) again. 

(\ref{eq:dtF2}) shows that our D8 is actually also charged under $F_2$, and thus that it is actually a D8/D6 bound state. 

In fact, we should mention that it also acts as a source for $H$. This should not come as a surprise: it comes from the fact that $B$ appears in the DBI brane action. The simplest way to see this phenomenon for us is to notice that $H$ in (\ref{eq:Hnice}) contains $F_0$. Since $F_0$ jumps across the D8, so does $H$, and its equation of motion now gets corrected to
\begin{equation}
	d(e^{7A-2 \phi} * H)- e^{7A} F_0 *F_2 = - x e^{7A-\phi} \Delta F_0 \delta \ .
\end{equation}
The localized term on the right hand side is exactly what one obtains by varying the DBI action $-\int_{S^2} e^{7A-\phi}\sqrt{\det(g+{\cal F})}$: the variation for a single D8 is $-e^{7A-\phi}\frac{{\cal F}}{\det(g+{\cal F})} \delta= -x e^{7A-\phi} \delta$. This was guaranteed to work: the equation of motion for $H$ was shown in \cite{koerber-tsimpis} to follow in general from supersymmetry even in presence of sources. (The CS term $\int C e^{\cal F}$ does not contribute, as remarked below \cite[(B.7)]{koerber-tsimpis}.)   

Yet another check one could perform is whether the D8 source is now BPS --- namely, whether the supersymmetry variation induced on its worldvolume theory can be canceled by an appropriate $\kappa$-symmetry transformation. This check is made simpler by the fact that brane calibrations are actually the same forms that appear in the bulk supersymmetry conditions (as first noticed in \cite{martucci-smyth} for compactifications to four dimensions). In our case, we see from \cite[Table 1]{lust-patalong-tsimpis} that the appropriate calibration for a space-filling brane is $e^{6A_4-\phi} {\rm Re} \phi^1_-$; for our AdS$_7$ case, we should pick in (\ref{eq:phipsi}) its part along $d\rho$. So our brane calibration is
\begin{equation}\label{eq:branecal}
	e^{7A-\phi}{\rm Im} \psi^1_+\ .
\end{equation}
The condition that a single brane should be BPS boils down to demanding that the pull-back of the form $e^{\cal F}{\rm Im} \psi^1_+$ equal the generalized volume form $\sqrt{{\rm det}(g+ {\cal F})}$ on the brane. Alternatively, this is equivalent to demanding that the pullback on the brane of 
\begin{equation}\label{eq:branecal2}
	e^{{\cal F}}{\rm Re} \psi^1_+ \ ,\qquad e^{{\cal F}} \psi^2_+
\end{equation}
vanish. We checked explicitly that this condition holds precisely if (\ref{eq:jump}) does. 

We should be a bit more careful, however, about what happens for multiple branes. In that case, (\ref{eq:branecal2}) become non-abelian, because they both contain the worldsheet field $f_2$. Satisfying this condition now requires ${\cal F}$ to be proportional to the identity, and this in turn requires that the D6-brane charge $\Delta n_2$ should be an integer multiple of $n_{\mathrm{D}8}= \Delta n_0$. In other words, a bunch of D8-branes should be made of magnetized branes which all have the same induced D6-brane charge.

Finally, in our analysis so far we have left out the case where $F_0$ is zero on one of the sides of the D8 stack, say the right side, so that $n_0'=0$. This time we cannot apply (\ref{eq:Bmassive}) on the right side of the D8. An expression for $B$ in this case will be given in (\ref{eq:Bmassless}) below. Imposing continuity of $B$ this time does not lead to (\ref{eq:jump}), but to a different condition in terms of the integration constants appearing in (\ref{eq:Bmassless}). However, the Bianchi identity for $F_2$ can still be applied on the left side of the D8, where $F_0 \neq 0$; this still leads to (\ref{eq:jump}). In other words, in this case we have (\ref{eq:jump}) plus an extra condition imposing continuity of $B$. This will be important in our example with two D8's in section \ref{sub:d8s}.

Let us summarize the results of this section. We have obtained that one can insert D8's in our setup, provided their position $r_{\mathrm{D}8}$ is such that the condition (\ref{eq:jump}) is satisfied. When $F_0$ is non-zero on both sides of the D8, this ensures that the Bianchi for $F_2$ is satisfied, and that $B$ is continuous. In the special case where $F_0=0$ on one side, continuity of $B$ has to be imposed independently. 


\subsection{Summary of this section} 
\label{sub:sum}

Supersymmetric solutions of the form AdS$_7\times M_3$ cannot exist in IIB. In IIA we have reduced the problem to solving the system of ODEs (\ref{eq:odeA}) (or (\ref{eq:oder})). Given a solution to this system, the flux is given by (\ref{eq:f0}), (\ref{eq:f2}) and (\ref{eq:H}), and the metric is given by (\ref{eq:metricfibr}) (or (\ref{eq:met-r})). This describes an $S^2$ fibration over a segment; the space is compact if the $S^2$ fiber shrinks at the endpoints of the segment, giving a topology $M_3=S^3$. This imposes the boundary conditions (\ref{eq:bc}) on the system (\ref{eq:oder}). D8-branes can be inserted along the $S^2$, at values $r=r_{\rm D8}$ that satisfy (\ref{eq:jump}).

We now turn to a numerical study of the system, which will show that nontrivial solutions do indeed exist. 


\section{Explicit solutions} 
\label{sec:exp}

We will now show some explicit AdS$_7\times M_3$ solutions, by solving the system (\ref{eq:oder}). We will start in section \ref{sub:massless} by looking briefly at the massless solution, which is in a sense unique; it has a D6-brane and an anti-D6 at the two poles. In section \ref{sub:nod8} we will switch on Romans mass, and we will obtain a solution with a D6 at one pole only. In section \ref{sub:d8s} we will then obtain regular solutions with D8-branes. 

\subsection{Warm-up: review of the $F_0=0$ solution} 
\label{sub:massless}

We will warm up by reviewing the solution one can get for $F_0=0$. 

As we remarked in section \ref{sub:geo}, in the massless case one can always lift to eleven-dimensional supergravity, and there we can only have AdS$_7\times S^4$ (or an orbifold thereof). The metric simply reads
\begin{equation}\label{eq:ads7s4}
	ds^2_{11}= R^2\left(ds^2_{{\rm AdS}_7} + \frac14 ds^2_{S^4}\right)\ ,
\end{equation}
being $R$ an overall radius. Let us now have a look at how this reduces to IIA. It is not obvious whether the reduction will preserve any supersymmetry; but, as we will now see, this can be arranged. 

To reduce, we have to choose an isometry. Since $S^4$ has Euler characteristic $\chi=2$, like any even-dimensional sphere, any vector field has at least two zeros, and so our reduction will have at least two points where the dilaton goes to zero; we expect some other strange feature at those two points, and as we will see this expectation is borne out. 

How should we choose the isometry? We can think about U(1) isometries on $S^d$ as rotations in $\rr^{d+1}$. The infinitesimal generator $v$ is an element of the Lie algebra $\mathfrak{so}(d+1)$, namely an antisymmetric  $(d+1)\times (d+1)$ matrix $v$. Moreover, two such elements $v_i$ that can be related by conjugation, $v_1 = O v_2 O^t$, for $O \in {\rm SO}(d+1)$, can be thought of as equivalent. Any antisymmetric matrix can be put in a canonical block-diagonal form where every block is of the form $\left(\begin{smallmatrix}0 & a \\ -a & 0 \end{smallmatrix}\right)$, with $a$ an angle. For even $d$, this implies that there is at least one zero eigenvalue, which corresponds to the fact that there is no vector field without zeros on the sphere. For $d=4$, we have two angles $a_1$ and $a_2$. Our solution can be reduced along any of these vector fields, but we also want the reduction to preserve some supersymmetry. The infinitesimal spinorial action of the vector field we just described is proportional to $a_1 \gamma_{12} + a_2 \gamma_{34}$. If we demand that this matrix annihilates at least one spinor $\chi$ (so that, at the finite level, $\chi$ is kept invariant), we get either $a_1=a_2$ or $a_1=-a_2$. 

To make things more concrete, let us introduce a coordinate system on $S^4$ adapted to the isometry we just found: 
\begin{equation}\label{eq:S4}
	ds^2_{S^4} = d \alpha^2 + \sin^2(\alpha) ds^2_{S^3}= d \alpha^2 + \sin^2(\alpha) \left(\frac14 d s^2_{S^2} + (dy+C_1)^2\right)\ ,\quad dC_1 = \frac12\vol_{S^2} \,
\end{equation}
with $ \alpha \in [0,\pi]$. We have written the $S^3$ metric as a Hopf fibration over $S^2$; the $1/4$ is introduced so that all spheres have unitary radius. The reduction will now proceed along the vector
\begin{equation}
	\del_y\ .
\end{equation} 
We can actually generalize this a bit by considering the orbifold $S^4/\zz_k$, where $\zz_k$ is taken to be a subgroup of the U(1) generated by $\del_y$. This is equivalent to multiplying the $(dy+C_1)^2$ term in (\ref{eq:S4}) by $\frac1{k^2}$. 

We can now reduce the eleven-dimensional metric (\ref{eq:ads7s4}), quotiented by the $\zz_k$ we just mentioned, using the string-frame reduction $ds^2_{11}= e^{-\frac23 \phi} ds^2_{10} + e^{\frac43 \phi} (dy + C_1)^2$. We obtain a metric of the form (\ref{eq:met73}), with
\begin{equation}\label{eq:massless}
	e^{2A} = R^2 e^{\frac23 \phi} = \frac {R^3}{2k} \sin(\alpha)\ ,\qquad 
	ds^2_{M_3}=\frac {R^3}{8k} \sin(\alpha) \left(d \alpha^2 + \frac14 \sin^2(\alpha) ds^2_{S^2}\right)\ . 
\end{equation}
We could now also reduce the Killing spinors on $S^4$, which are given in appendix \ref{app:kill} in our coordinates. There are indeed two of them which can be reduced, confirming our earlier arguments. This would allow us to compute directly the $\psi^{1,2}$. We will instead proceed by using the equations we derived in section \ref{sec:gen}. It is actually more convenient, in this case, to work directly with the system (\ref{eq:odeA}), that can be more easily solved explicitly: 
\begin{equation}\label{eq:xmassless}
	x=\sqrt{1-e^{4(A-A_0)}} \ ,\qquad \phi=3A - \phi_0\ 
\end{equation}
where $A_0$ and $\phi_0$ are two integration constants. This can be seen to be the same as (\ref{eq:massless}) by taking 
\begin{equation}
	x= \cos(\alpha) \ ,\qquad A_0 = \frac12 \log \left(\frac{R^3}{2k}\right) 
	\ ,\qquad \phi_0 = 3 \log R \ .
\end{equation}
The fluxes can now be computed from (\ref{eq:f2}) and (\ref{eq:H}):
\begin{equation}\label{eq:fluxmassless}
	F_2= -\frac12 k {\rm vol}_{S^2} \ ,\qquad 
	H = -\frac3{32}\frac{R^3}{k} \sin^3(\alpha) d \alpha \wedge {\rm vol}_{S^2}\ ;
\end{equation}
the $B$ field then can be written as 
\begin{equation}\label{eq:Bmassless}
	B_2= \frac3{32} \frac{R^3}k \left( x - \frac{x^3}3 \right) {\rm vol}_{S^2}+ b
\end{equation}
where again $b$ is a closed two-form. The simple result for $F_2$ in (\ref{eq:fluxmassless}) could be expected from the fact that the metric (\ref{eq:S4}) is an $S^1$ fibration over $S^2$ with Chern class $c_1=-k$.

However, (\ref{eq:massless}) might appear problematic for two reasons. First of all, the warping function goes to zero at the two poles $\alpha=0$, $\alpha=\pi$.\footnote{The warping function also goes to zero at the equator of the AdS$_6\times S^4$ solution \cite{brandhuber-oz}, recently shown \cite{passias} to be the only AdS$_6$ solution in massive IIA. This solution can also be T-dualized, without breaking supersymmetry, both using its non-abelian and the more usual abelian isometries \cite{lozano-colgain-rodriguezgomez-sfetsos}, differently from what we saw for AdS$_7$ in section \ref{sub:metric}.} Second, $ds^2_{M_3}$ would be singular at the poles even if it were not multiplied by an overall factor $e^{2A}=\frac {R^3}{2k} \sin(\alpha)$, because of the $1/4$ in front of $ds^2_{S^2}$. Indeed, when we expand it around, say, $\alpha=0$, we find $d\alpha^2+ \frac{\alpha^2}4 ds^2_{S^2}$; this would be regular without the $1/4$, but as it stands it has a conical singularity. 

However, these singularities at the poles have the behavior one expects near a D6. Near the north pole $\alpha=0$, $ds^2_{M_3}$ in (\ref{eq:massless}) looks like $ds^2_{M_3} \sim \alpha \left(d \alpha^2 + \frac14 \alpha^2 ds^2_{S^2}\right)$. In terms of the $r$ variable we used in (\ref{eq:met-r}), this looks like 
\begin{equation}\label{eq:nearD6}
	ds^2_{M_3} \sim d r^2 + \left(\frac34 r\right)^2 ds^2_{S^2}\ .
\end{equation}
Near the ordinary flat-space D6-brane metric, $ds^2_{M_3} \sim \rho^{-1/2} (d \rho^2 + \rho^2 ds^2_{S^2})$, which also looks like (\ref{eq:nearD6}) with $r= \frac43 \rho^{3/4}$.

The presence of D6's could actually be inferred more directly. First of all, we know that D6-branes result from loci where the size of the eleventh dimension goes to zero; this indeed happens at the two poles. Moreover, from the expression of $F_2$ in (\ref{eq:fluxmassless}), the integral of $F_2$ over the $S^2$ is constant and equal to $-2\pi k$. We can take the $S^2$ close to the north or the south pole, where it signals the presence of D6-brane charge. More precisely, there are $k$ anti-D6-branes at the north pole and $k$ D6-branes at the south pole. 

One crucial difference with the usual D6 behavior, however, is the presence of the NS three-form $H$. From (\ref{eq:fluxmassless}) we see that it does not vanish near the D6. Rather, it diverges: near the anti-D6 at $r=r_{\rm N}=0$,\footnote{It is interesting to ask what happens in the Minkowski limit. From (\ref{eq:Hnice}) we see that $H=-6e^{-A} \vol_3$; taking $R\to \infty$, $e^{-A}$ tends to zero except than in a region $\alpha \ll R^{-1/3}$, which gets smaller and smaller in the limit.}
\begin{equation}\label{eq:divH}
	H \sim r^{-1/3} \vol_3\ .
\end{equation} 
This can also be inferred directly from eleven-dimensional supergravity, using the reduction formula $G_4= e^{\phi/3} H \wedge e^{11}$. Since $\phi\sim r$, the three-form energy density diverges as $e^{-2\phi} H^2 \sim (r_{\rm N}-r)^{-8/3}$. We should remember, in any case, that this solution is non-singular in eleven dimensions; the diverging behavior in (\ref{eq:divH}) is cured by M-theory, just like the divergence of the curvature of (\ref{eq:nearD6}) is.

The simultaneous presence of D6's and anti-D6's in a BPS solution might look unsettling at first, since in flat space they cannot be BPS together. It is true that the conditions imposed on the supersymmetry parameters $\epsilon_i$ by a D6 and by an anti-D6 brane are incompatible. But in flat space the $\epsilon_i$ are constant, while in our present case they are not. The condition changes from the north pole to the south pole; so much so that an anti-D6 is BPS at the north pole, and a D6 is BPS at the south pole. Although we have not been working explicitly with spinors in this paper, but rather with forms, we can see this by performing a brane probe analysis in the language of calibrations, as we did for D8-branes at the end of section \ref{sub:d8}. The relevant polyform is again (\ref{eq:branecal}); for a D6 we should use its zero-form part, which from (\ref{eq:psigen}) is simply $\cos(\theta_1) \sin(\psi)= x$. For a D6 or anti-D6, this should be equal to plus or minus the internal volume form of the D6, which is $\pm 1$; this happens precisely at the north and south pole.

\begin{figure}[ht]
	\centering
		\includegraphics[scale=1]{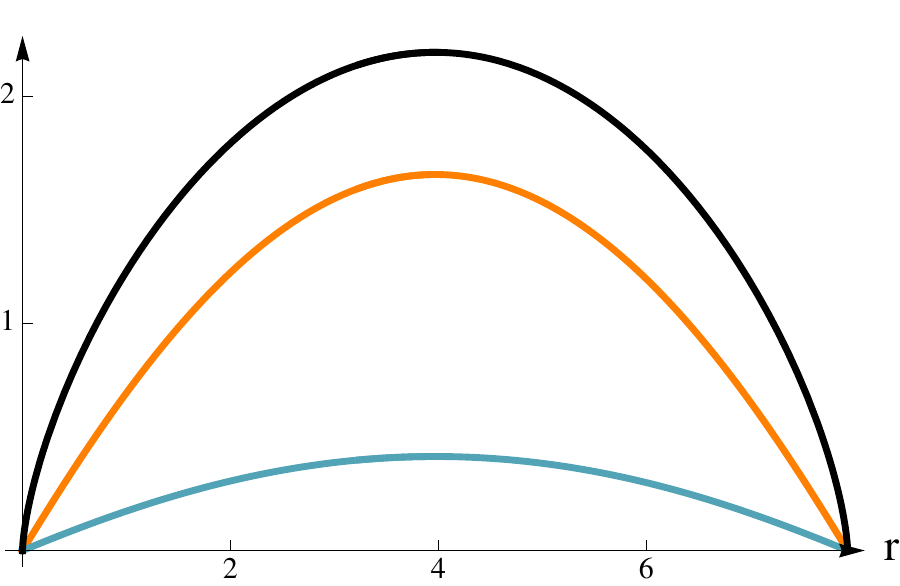}
	\caption{Massless solution in IIA. We show here the radius of the $S^2$ (orange), the warping factor $e^{2A}$ (black; multiplied by a factor $1/20$), and the string coupling $e^\phi$ (green; multiplied by a factor $5$). We see that the warping goes to zero at the two poles. The angular coefficient of the orange line can be seen to be $3/4$ as in (\ref{eq:nearD6}). The two singularities are due to $k$ D6 and $k$ anti-D6 (in this picture, $k=20$).}
	\label{fig:massless}
\end{figure}

In figure \ref{fig:massless} we show some parameters for the solution as a function of the $r$ defined in (\ref{eq:met-r}), for uniformity with latter cases. We also show the radius of the transverse sphere, which near the poles has the angular coefficient $3/4$ of (\ref{eq:nearD6}).

We have obtained this massless IIA solution by reducing the M-theory solution ${\rm AdS}_7 \times S^4/\zz_k$, but other orbifolds would be possible as well. One could for example have quotiented by the $\hat D_{k-2}$ groups, which would have resulted in IIA in an orientifold by the action of the antipodal map on the $S^2$. The transverse $S^2$ would have been replaced by an $\rr\pp^2$; at the poles we would have had O6's together with the $k$ D6's/anti-D6's of the $A_k$ case. 

We will see in section \ref{sub:d8s} solutions with $F_0 \neq 0$ and without any D6-branes. But we will at first try in the next subsection to introduce $F_0$ without any D8-branes. 


\subsection{Massive solution without D8-branes} 
\label{sub:nod8}

In section \ref{sub:massless} we reviewed the only solution for $F_0=0$, related to AdS$_7\times S^4$ by dimensional reduction; it has a D6 and an anti-D6 at the poles of $M_3\cong S^3$. 

We now start looking at what happens in presence of a non-zero Romans mass, $F_0\neq0$. We saw in section \ref{sub:loc} that in this case it is possible for the poles to be regular points. It remains to be seen whether those boundary conditions can be joined by a solution of the system (\ref{eq:oder}).  

\begin{figure}[ht]
\centering	
	\subfigure[\label{fig:nod8-a}]{\includegraphics[scale=.9]{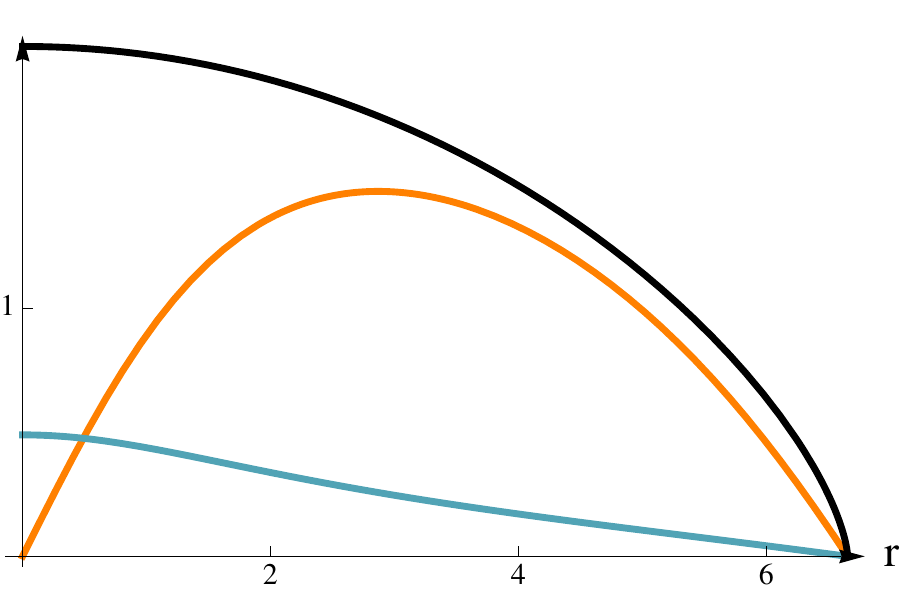}}
	\hspace{.4cm}
	\subfigure[\label{fig:nod8-b}]{\includegraphics[scale=.7]{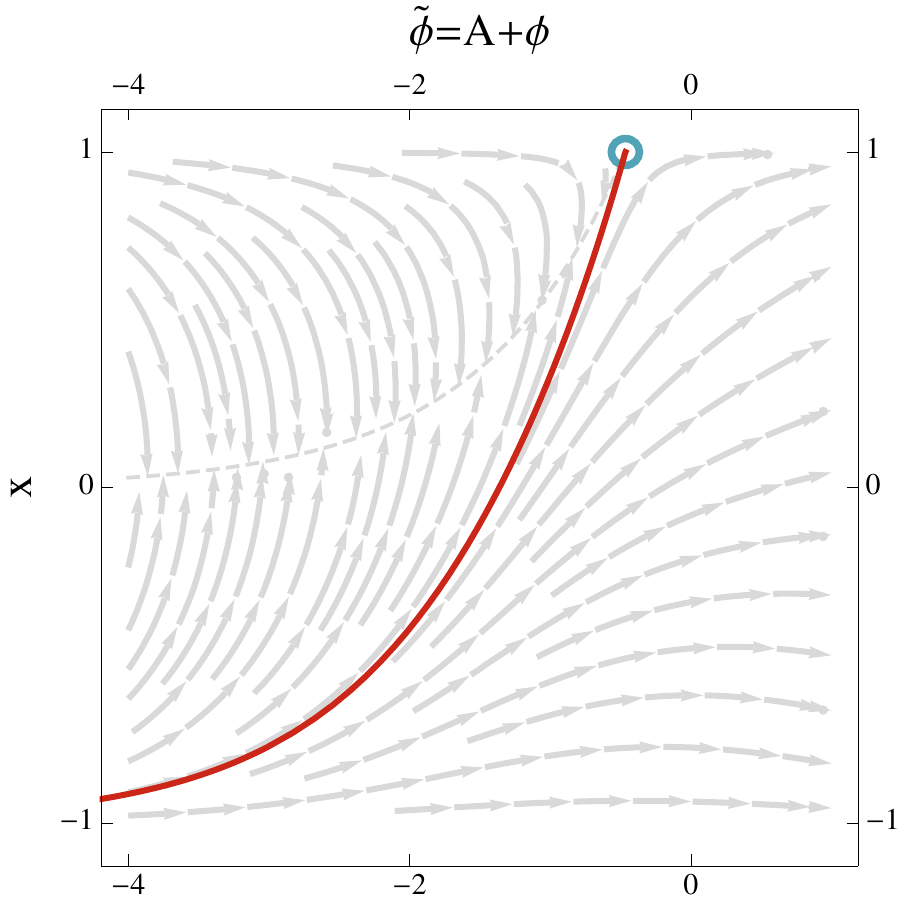}}
	\caption{Solution for $F_0=40/2\pi$. We imposed regularity at the north pole, and evolved towards positive $r$. In \subref{fig:nod8-a} we again plot the radius of the $S^2$ (orange), the warping factor $e^{2A}$ (black; multiplied by a factor $1/20$), and the string coupling $e^\phi$ (green; multiplied by a factor $5$). With increasing $r$, the plot gets more and more similar to the one for the massless case in figure \ref{fig:massless}. There is a stack of D6's at the south pole (in this picture, $k=112$ of them), as in the massless case, although this time it also has a diverging NS three-form $H$. Notice that the size of the $S^2$ goes linearly near both poles, but with angular coefficients $1$ near the north pole (appropriate for a regular point) and $3/4$ for the south pole (appropriate for a D6, as seen in (\ref{eq:nearD6})). In \subref{fig:nod8-b}, we see the path described by the solution in the $\{ A+ \phi, x\}$ plane, overlaid to the vector field shown in figure \ref{fig:stream}.}
	\label{fig:nod8}
\end{figure}

We can for example impose the boundary condition (\ref{eq:bc}) at $r=r_{\rm N}$, and evolve numerically towards positive $r$ using (\ref{eq:oder}). The procedure is standard: we use the approximate power-series solution (\ref{eq:expN}) from $r=r_{\rm N}=0$ to a very small $r$, and then use the values of $A$, $\phi$, $x$ thus found as boundary conditions for a numerical evolution of (\ref{eq:oder}). One example of solution is shown in figure \ref{fig:nod8-a}. It stops at a finite value of $r$, where it resembles there the south pole behavior of the massless case in figure \ref{fig:massless}; for example, $e^A$ goes to zero at the right extremum.

This is actually easy to understand already from the system, both in (\ref{eq:odeA}) and in (\ref{eq:oder}). As $A$ and $\phi$ get negative, they suppress the terms containing $F_0$, and the system tends to the one for the massless case. 

An alternative, and perhaps more intuitive, understanding can be found using the form (\ref{eq:odeA}) of the system, which we drew in figure \ref{fig:stream} as a vector field flow on the space $\{ A+\phi,x\}$. The green circle in that figure represents the point $\{A+\phi=\log(4/F_0),x=1\}$, which is the appropriate boundary condition for the north pole in (\ref{eq:bc}). In that figure the `time' variable is $A$. From (\ref{eq:expN}), we see that $A$ has a local maximum at $r=r_{\rm N}$. So the stream in figure \ref{fig:stream} has to be followed backwards, starting from the green circle at the top. We can see that the integral curve asymptotically approaches $x=-1$, but does not get there in finite `time'; in other words, $A\to -\infty$. The flow corresponding to the solution in figure \ref{fig:nod8-a} is shown in figure \ref{fig:nod8-b}.

In the massless case, we saw in section \ref{sub:massless} that the singularities at the poles are actually D6-branes. In this case too we have D6's at the south pole. This is confirmed by considering the integral of $F_2$ along a sphere $S^2$ in the limit where it reaches the south pole: it gives a non-zero number. By tuning $A_0^+$, this can be arranged to be $2\pi$ times an integer $k$, where $k$ is the number of D6-branes at the south pole. The presence of these D6-branes without any anti-D6 is not incompatible with the Bianchi identity $dF_2 - H F_0 = k\delta_{D6}$, because integrating it gives $-F_0 \int H = k$. In other words, the flux lines of the D6's are absorbed by $H$-flux, as is often the case for flux compactifications. Notice also that these D6's are calibrated; the computation runs along similar lines as the one we presented for the massless solution in section \ref{sub:massless}.

To be more precise, the singularity is not the usual D6 singularity, in that there is also a NS three-form $H$ diverging as in (\ref{eq:divH}). 
This is consistent with the prediction in \cite[Eq.~(4.15)]{gautason-junghans-zagermann} (given there in Einstein frame), and in general with the analysis of \cite{blaback-danielsson-junghans-vanriet-wrase-zagermann,blaback-danielsson-junghans-vanriet-wrase-zagermann-2}, which found that it is problematic to have ordinary D6-brane behavior in a massive AdS$_7\times S^3$ setup precisely like the one we are considering here. (In the language of \cite{blaback-danielsson-junghans-vanriet-wrase-zagermann}, the parameter $\alpha$ of our solution goes to a negative constant; this enables the solution to exist and to evade the global no-go they found, but at the cost of the diverging $H$ in (\ref{eq:divH}), \cite[Eq.~(4.15)]{gautason-junghans-zagermann}.) More precisely, the asymptotic behavior we find is the one discovered in \cite[Eq.~(3.4)]{blaback-danielsson-junghans-vanriet-wrase-zagermann-2}.

Thus the singularity at the south pole in figure \ref{fig:nod8} is the same we found in the massless case we saw in section \ref{sub:massless}. In that case, the singularity is cured by M-theory. In the present case, the non-vanishing Romans mass prevents us from doing that. However, we still think it should be interpreted as the appropriate response to a D6; for this reason we think it is a physical solution. 

So far we have examined what happens when we impose that the north pole is regular. It is also possible to have a D6 and anti-D6 singularity at both poles, as in the previous section, or an O6 at one of the poles (keeping D6's at the other pole). Roughly speaking, this corresponds to a trajectory similar to the one in figure \ref{fig:nod8-b}, in which one ``misses'' the green circle to the left or to the right, respectively. As we have seen, the D6 solution is very similar to the massless one. The O6 solutions also turn out to be very similar to their massless counterpart:\footnote{In the different setup of \cite{saracco-t}, an O6 in presence of $F_0$ gets modified in such a way that its singularity disappears. This does not happen here.} near the pole, their asymptotics is $e^A\sim r^{-1/5}$, $e^\phi\sim r^{-3/5}$, $x\sim 1 - r^{4/5}$. This leads to  the same asymptotics for the metric as in the massless O6 solution near the critical radius $\rho_0 = g_s l_s$. Once again, however, in the massive case we have a diverging NS three-form; this time $H\sim r^{-3/5} {\rm vol}_3$. Finally, in such a case the $S^2$ is replaced by an $\rr\pp^2$ because of the orientifold action. 


\subsection{Regular massive solution with D8-branes} 
\label{sub:d8s}

We will now examine what happens in presence of D8-branes. 

The first possibility that comes to mind is to put all of them together in a single stack. The idea is the following. We once again use the power-series expansion (\ref{eq:expN}) from $r=r_{\rm N}=0$ to a small $r$, and use the resulting values of $A$, $\phi$ and $x$ as boundary conditions for a numerical evolution of (\ref{eq:oder}). This time, however, we should stop the evolution at a value of $r$ where (\ref{eq:jump}) is satisfied. At this point $F_0$ will change, and (\ref{eq:oder}) will change as well. Generically, the evolution on the other side of the D8 will lead to a D6 or an O6 singularity, as discussed in section \ref{sub:nod8}. However, if $F_0$ is negative, according to (\ref{eq:bc}), the point $\{x=-1,\,e^{A+\phi}=-\frac4{F_0}\}$ leads to a regular South Pole. Fortunately, our solution still has a free parameter, namely $A_0^+=A(r_{\rm N})$. By fine-tuning this parameter, we can try to reach $\{x=-1,\,e^{A+\phi}=-\frac4{F_0}\}$ and obtain a regular solution. 

Alternatively, after stopping the evolution from the North Pole to the D8, one can look for a similar solution starting from the South Pole, and then match the two --- in the sense that one should make sure that $A$, $\phi$, and $x$ are continuous. One combination of them, namely $q$, will already match by construction. It is then enough to match two variables, say $A$ and $x$; this can be done by adjusting $A_0^+$ and $A_0^-$. 

Naively, however, we face a problem when we try to choose the flux parameters on the two sides of the D8's. We concluded in (\ref{eq:b0}) that near the poles we should have $b=0$; this seems to imply, via (\ref{eq:bn2}), that $n_2=0$ on both sides of the D8. (\ref{eq:jump}) would then lead to $q=0$ on the D8, which can only be true at the poles $x=\pm1$. 

This confusion is easily cleared once we remember that $B$ can undergo a large gauge transformation that shifts it by $k \pi {\rm vol}_{S^2}$, as we explained towards the end of section \ref{sub:loc}. We saw there that we can keep track of this by introducing the variable $\hat b$ in (\ref{eq:hatb}). We now simply have to make sure that $\hat b$ winds an integer amount of times $N$ around the fundamental domain $[0,\pi]$; this can be interpreted as the presence of $N$ large gauge transformations, or as the presence of a non-zero quantized flux $N= \frac1{4\pi^2}\int H$. 

We still face one last apparent problem. It might seem that making sure that $\hat b$ winds an integer amount of times requires a further fine-tuning on the solution; this we cannot afford, since we have already used both our free parameters $A_0^\pm$ to make sure all the variables are continuous, and that the poles are regular. 

Fortunately, such an extra fine-tuning is in fact not necessary. Let us call $(n_0,n_2)$ the flux parameters before the D8, and $(n_0',n_2')$ after it. For simplicity let us also assume $n_2'=0$, so that no large gauge transformations are needed on that side. As we remarked at the end of section \ref{sub:d8}, $\Delta n_2 = n_2'-n_2 = -n_2$ should be an integer multiple of $\Delta n_0 = n_0'-n_0= n_{\rm D8}$: $\Delta n_2= \mu \Delta n_0$, $\mu \in \zz$. To take care of flux quantization, it is enough to also demand that $n_2= N n_0$ for $N$ integer. Indeed, from (\ref{eq:F0n0}), (\ref{eq:bn2}), (\ref{eq:hatb}), we see that in that case at the North Pole we get $\hat b=- \pi N$; since this is an integer multiple of $\pi$, it can be brought to zero by using large gauge transformations. Together, the conditions we have imposed determine $n_0' = n_0 \left(1-\frac N \mu\right)$. 

\begin{figure}[ht]
\centering	
	\subfigure[\label{fig:1d8-a}]{\includegraphics[scale=.9]{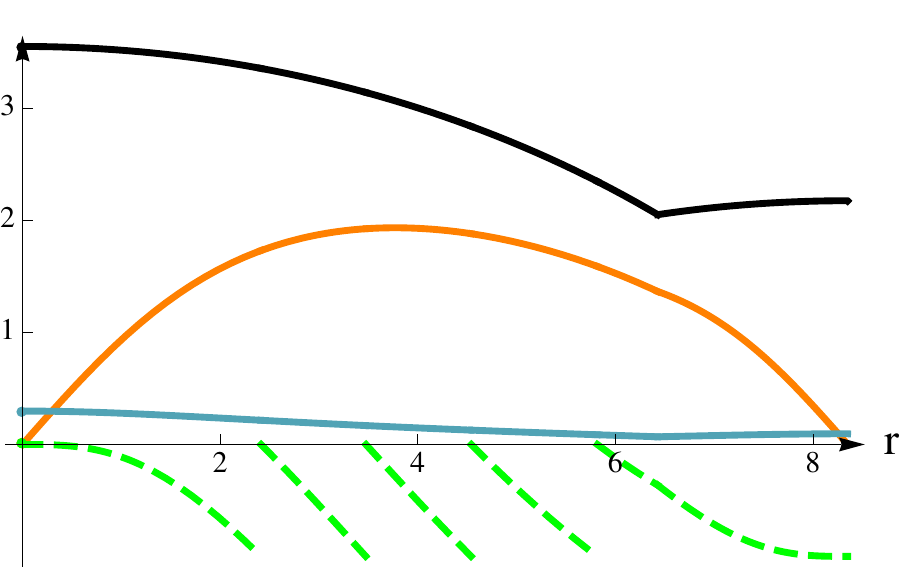}}
	\hspace{.4cm}
	\subfigure[\label{fig:1d8-b}]{\includegraphics[scale=.7]{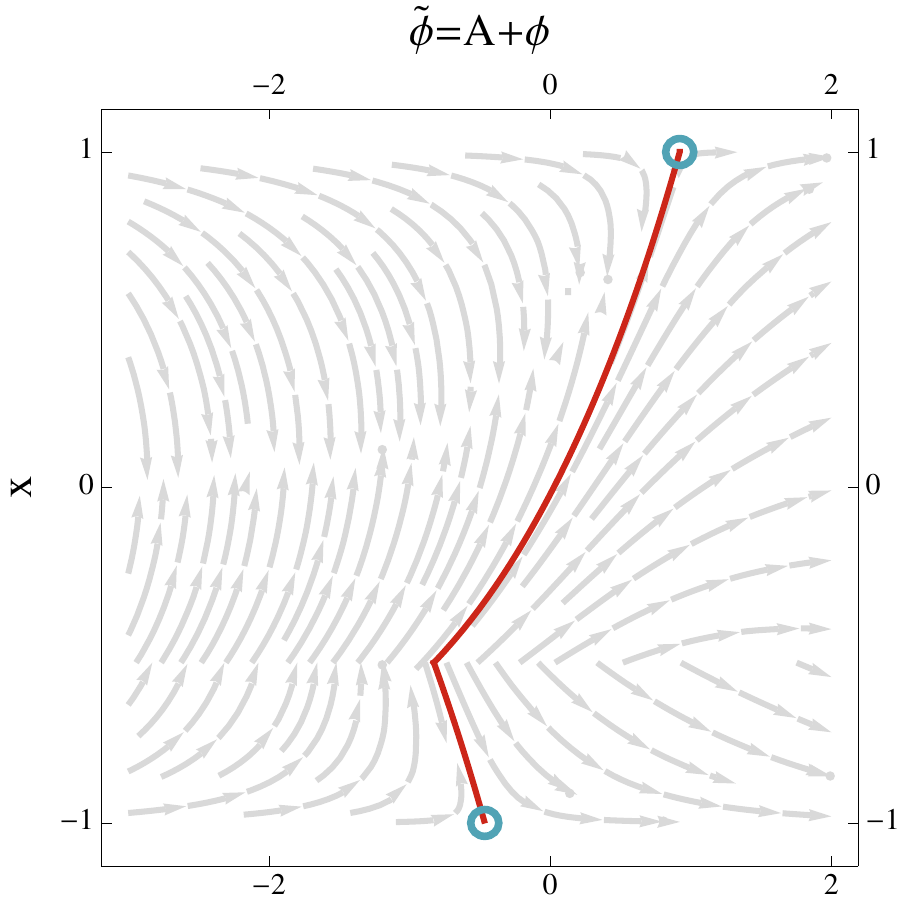}}
	\caption{Regular solution with one D8 stack. Its position can be seen in the graph as the value of $r$ where the derivatives of the functions jump; it is fixed by (\ref{eq:jump}). In \subref{fig:1d8-a} we again plot the radius of the $S^2$ (orange), the warping factor $e^{2A}$ (black; rescaled by a factor $1/20$), and the string coupling $e^\phi$. We also plot $\frac1{\pi}\hat b(r)=\frac1{4\pi^2}\int_{S^2_r} B_2$ (dashed, light green); to guide the eye, we have periodically identified it as described in section \ref{sub:loc}. (The apparent discontinuities are an artifact of the identification.)  The fact that it starts and ends at $\hat b=0$ is in compliance with flux quantization for $H$; we have $\frac1{4\pi^2}\int H= -5$. The flux parameters are $\{n_0,n_2\}=\{10,-50\}$ on the left (namely, near the north pole), $\{-40,	0\}$ on the right (near the south pole). In \subref{fig:2d8-b}, we see the path described by the solution in the $\{ A+ \phi, x\}$ plane, overlaid to the relevant vector field, that this time changes with $n_0$.}
	\label{fig:1d8}
\end{figure}

All this gives a strategy to obtain solutions with one D8 stack. We show one concrete example in figure \ref{fig:1d8}. One might find it intuitively strange that the D8-branes are not ``slipping'' towards the South Pole. The branes back-react on the geometry, bending the $S^3$, much as a rubber band on a balloon. This by itself, however, would not be enough to prevent them from slipping. Rather, we also have to take into account the Wess--Zumino term in the brane action. This term, which takes into account the interaction of the branes with the flux, balances with the gravitational DBI term to stabilize the D8's. The formal check of this is that the branes are calibrated, something we have already seen in section \ref{sub:d8} (see discussion around (\ref{eq:branecal}), (\ref{eq:branecal2})). The D8 stack is made of $n_{\rm D8}=50$ D8-branes; each of these D8's has worldsheet flux $f_2$ such that $\int_{S^2} f_2 = -2 \pi$, which means that it has an effective D6-brane charge equal to $-1$. A single D8/D6 bound state probe with this charge is calibrated exactly at $r=r_{\rm D8}$, and thus will not slip to the South Pole.
The solution can perhaps be thought of as arising from the one in figure \ref{fig:nod8} via some version of Myers' effect.\footnote{We thank I.~Bena, S.~Kuperstein, T.~Van Riet and M.~Zagermann for very useful conversations about this point and about the existence of solutions with a single D8. These solutions are consistent with the analysis in \cite{junghans-schmidt-zagermann}.} 

We can also look for a configuration with two stacks of D8-branes, again with regular poles. The easiest thing to attempt is a symmetric configuration where the two stacks have the same number of D8's, with opposite D6 charge. As for the solution with one D8, (\ref{eq:bc}) implies $F_0$ at the north pole and negative $F_0$ at the south pole. For our symmetric configuration, these two values will be opposite, and there will be a central region between the two D8 stacks where $F_0=0$. 

\begin{figure}[ht]
\centering	
	\subfigure[\label{fig:2d8-a}]{\includegraphics[scale=.9]{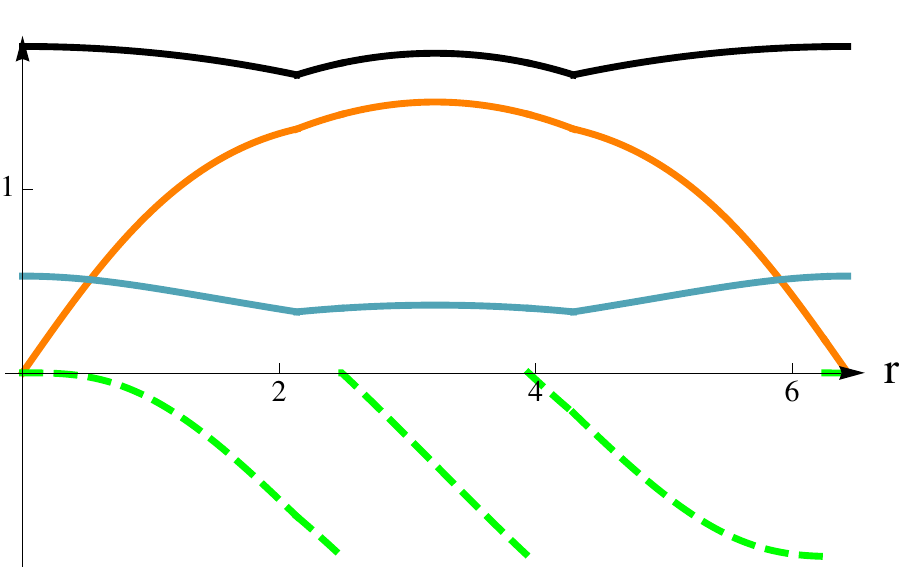}}
	\hspace{.4cm}
	\subfigure[\label{fig:2d8-b}]{\includegraphics[scale=.7]{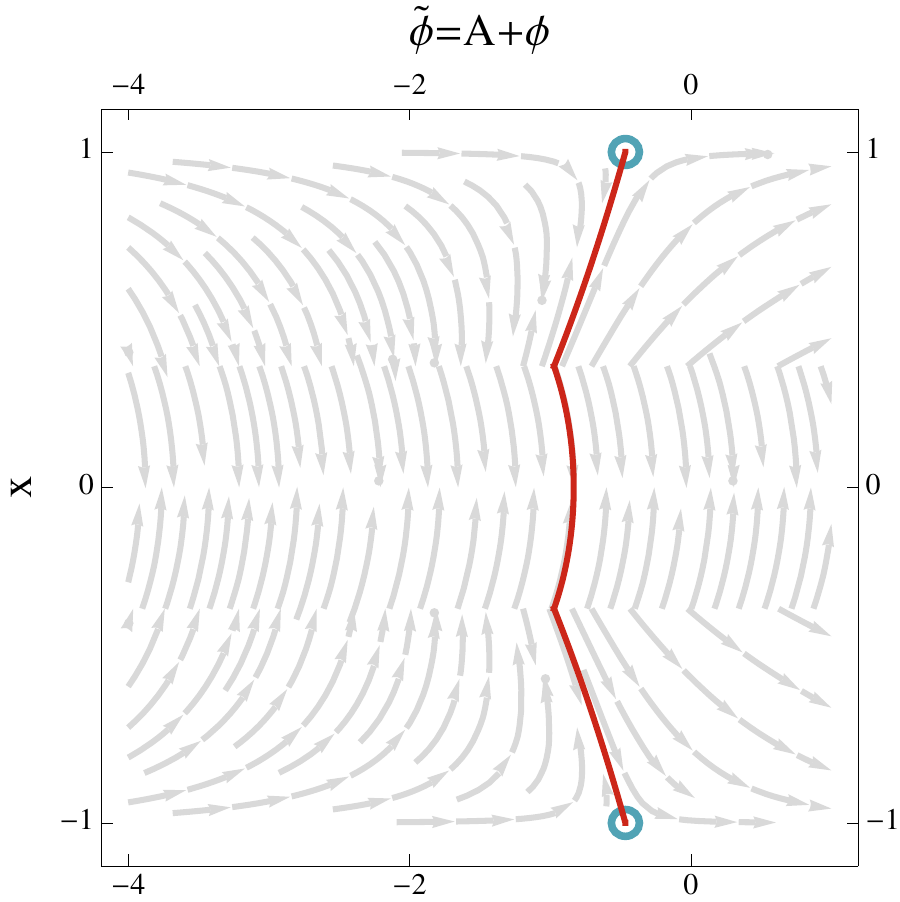}}
	\caption{Regular solution with two D8 stacks. As in figure \ref{fig:1d8}, their positions are the two values of $r$ where the derivatives of the functions jump. In \subref{fig:2d8-a} we again plot the radius of the $S^2$ (orange), the warping factor $e^{2A}$ (black; rescaled by a factor $1/20$), and the string coupling $e^\phi$ (green; rescaled by a factor $5$), and $\hat b$ (as in figure \ref{fig:1d8}; this time $\frac1{4\pi^2}\int H= -3$). The flux parameters are: $\{n_0,n_2\}=\{40,0\}$ on the left (namely, near the north pole); $\{0,-40\}$ in the middle; $\{-40,0\}$ on the right (near the south pole). The region in the middle thus has $F_0=0$; it is indeed very similar to the massless case of figure \ref{fig:massless}. In \subref{fig:2d8-b}, we see the path described by the solution in the $\{ A+ \phi, x\}$ plane, overlaid to the relevant vector field, that again changes with $n_0$.}
	\label{fig:2d8}
\end{figure}

We show one such solution in figure \ref{fig:2d8}. As for the previous solution with one D8, we have started from the North Pole and South Pole; now, however, we did not try to match these two solutions directly, but we inserted a massless region in between. From the northern solutions, again we found at which value of $r=r_{\rm D8}$ it satisfies (\ref{eq:jump}). We then stopped the evolution of the system there, evaluated $A$, $\phi$, $x$ at $r_{\rm D8}$, and used them as a boundary condition for the evolution of (\ref{eq:oder}), now with $F_0=0$. Now we matched this solution to the southern one; namely, we found at which values of $r=r_{\rm D8'}$ their $A$, $\phi$ and $x$ matched. This requires translating the southern solution in $r$ by an appropriate amount, and picking $A_0^-=A_0^+$. Given the symmetry of our configuration, this is not surprising: the southern solution is related to the northern one under (\ref{eq:F0-F0}). Moreover, matching a region with $F_0 \neq 0$ to the massless one means imposing an extra condition, namely the continuity of $B$ in $r_{\rm D8}$, as we mentioned at the end of \ref{sub:d8}.

The parameter $A_0^+=A_0^-=A_0$ would at this point be still free. However, one still has to impose flux quantization for $H$. As we recalled above, this is equivalent to requiring that the periodic variable $\hat b$ starts and ends at zero. Unlike the case with one D8 above, this time we do need a fine-tuning to achieve this, since the expression for $B$ is not simply controlled by the massive expression (\ref{eq:Bmassive}). Fortunately we can use the parameter $A_0$ for this purpose. The solution in the end has no moduli.

As for the solution with one D8 stack we saw earlier, in this case too the D8-branes are not ``slipping'' towards the North and South Pole because of their interaction with the RR flux: each of the two stacks is calibrated. In this case, intuitively this interaction can be understood as the mutual electric attraction between the two D8 stacks, which indeed have opposite charge under $F_2$; the balance between this attraction and the ``elastic'' DBI term is what stabilizes the branes. 

Let us also remark that for both solutions (the one with one D8 stack, and the one with two) it is easy to make sure, by taking the flux integers to be large enough, that the curvature and the string coupling $e^\phi$ are as small as one wishes, so that we remain in the supergravity regime of string theory. In figures 
\ref{fig:1d8} and \ref{fig:2d8} they are already rather small (moreover, in the figure we use some rescalings for visualization purposes). 

Thus we have found regular solutions, with one or two stacks of D8-branes. It is now in principle possible to go on, and to add more D8's. We have found examples with four D8 stacks, which we are not showing here. We expect that generalizations with an arbitrary number of stacks should exist, especially if there is a link with the brane configurations in \cite{brunner-karch,hanany-zaffaroni-6d}. Another possibility that might also be realized is having an O8-plane at the equator of the $S^3$.



\section*{Acknowledgments}
We would like to thank I.~Bena, F.~Gautason, S.~Kuperstein, D.~Martelli, A.~Passias, D.~Tsimpis, T.~Van Riet, A.~Zaffaroni and M.~Zagermann for interesting discussions. 
F.A.~is grateful to the Graduiertenkolleg GRK 1463 ``Analysis, Geometry and String Theory'' for support. The work of M.F.~was partially supported by the ERC Advanced Grant ``SyDuGraM'', by IISN-Belgium (convention 4.4514.08) and by the ``Communaut\'e Fran\c{c}aise de Belgique" through the ARC program. M.F.~ is a Research Fellow of the Belgian FNRS-FRS. D.R.~and A.T.~are supported in part by INFN, by the MIUR-FIRB grant RBFR10QS5J ``String Theory and Fundamental Interactions'', and by the MIUR-PRIN contract 2009-KHZKRX. The research of A.T.~is also supported by the  European Research Council under the European Union's Seventh Framework Program (FP/2007-2013) -- ERC Grant Agreement n. 307286 (XD-STRING). 

\appendix

\section{Supercharges} 
\label{app:gamma}

At the beginning of section \ref{sub:73} we reviewed an old argument that shows how a solution of the form AdS$_7\times M_3$ can also be viewed as a solution of the type Mink$_6\times M_4$. In this appendix we show how the AdS$_7\times M_3$ supercharges get translated in the Mink$_6\times M_4$ framework. 

A decomposition of gamma matrices appropriate to six-dimensional compactifications reads 
\begin{equation}\label{eq:gamma64}
	\gamma^{(6+4)}_\mu = e^{A_4} \gamma^{(6)}_\mu \otimes 1 \ ,\qquad
	\gamma^{(6+4)}_{m+5} = \gamma^{(6)} \otimes \gamma^{(4)}_m \ .
\end{equation}
Here $\gamma^{(6)}_\mu$, $\mu=0,\ldots,5$, are a basis of six-dimensional gamma matrices, while $\gamma^{(4)}_m$, $m=1,\ldots,4$ are a basis of four-dimensional gamma matrices. For a supersymmetric Mink$_6 \times M_4$ solution, the supersymmetry parameters can be taken to be
\begin{equation}\label{eq:eps64}
\begin{split}
	\epsilon_1^{(6+4)}= \zeta^0_+ \otimes \eta^1_+ + \zeta^{0\,c}_+ \otimes \eta^{1\,c}_+ \ , \\
	\epsilon_2^{(6+4)}= \zeta^0_+ \otimes \eta^2_\mp + \zeta^{0\,c}_+ \otimes \eta^{2\,c}_\mp\ ,
\end{split}
\end{equation}
where $\zeta_+$ is a constant spinor; ${}_\mp$ denotes the chirality, and ${}^c$ Majorana conjugation both in six and four dimensions. Supersymmetry implies that the norms of the internal spinors satisfy $|| \eta^1 ||^2 \pm || \eta^2 ||^2= c_{\pm} e^{\pm A_4}$, where $c_\pm$ are constant. 
 
On the other hand, for seven-dimensional compactifications a possible gamma matrix decomposition reads
\begin{equation}\label{eq:gamma73}
	\begin{split}
		&\gamma^{(7+3)}_\mu = e^{A_3} \gamma^{(7)}_\mu \otimes 1 \otimes \sigma_2 \ ,\\
		&\gamma^{(7+3)}_{i+6} = 1 \otimes \sigma_i \otimes \sigma_1 \ .
	\end{split} 
\end{equation}
This time $\gamma^{(7)}_\mu$, $\mu=0,\ldots,6$, are a basis of seven-dimensional gamma matrices, and $\sigma_i$, $i=1,2,3$, are a basis of gamma matrices in three dimensions (which in flat indices can be taken to be the Pauli matrices). For a supersymmetric solution of the form AdS$_7\times M_3$, the supersymmetry parameters are now of the form 
\begin{equation}\label{eq:eps73}
\begin{split}
		&\epsilon_1^{(7+3)} = (\zeta\otimes \chi_1 + \zeta^c \otimes \chi_1^c) \otimes v_+\ , \\
		&\epsilon_2^{(7+3)} = (\zeta\otimes \chi_2 \mp \zeta^c \otimes \chi_2^c) \otimes v_\mp \ .
\end{split}
\end{equation}
Here, $\chi_{1,2}$ are spinors on $M_3$, with $\chi_{1,2}^c \equiv B_3 \chi_{1,2}^*$ their Majorana conjugates; a possible choice of $B_3$ is $B_3=\sigma_2$. $\zeta$ is a spinor on AdS$_7$, and $\zeta^c \equiv B_7 \zeta^*$ is its Majorana conjugate; there exists a choice of $B_7$ which is real and satisfies $B_7 \gamma_\mu = \gamma_\mu^* B_7$. (It also obeys $B_7 B_7^*=-1$, which is the famous statement that one cannot impose the Majorana condition in seven Lorentzian dimensions.) The ten-dimensional conjugation matrix can then be taken to be $B_{10}= B_7 \otimes B_3 \otimes \sigma_3$; the last factor in (\ref{eq:eps73}), 
$v_\pm$, are then spinors chosen in such a way as to give the $\epsilon_i^{(7+3)}$ the correct chirality, and to make them Majorana; with the above choice of $B_{10}$,  $v_+=\frac1{\sqrt2}{1 \choose -1}$, $v_-=\frac1{\sqrt2}{1\choose 1}$. The minus sign (for the IIA case) in front of the term $\zeta^c \otimes \chi_2^c$ in (\ref{eq:eps73}) is due to the fact that, both in seven Lorentzian and three Euclidean dimensions, conjugation does not square to one: $(\zeta^c)^c= - \zeta$, $(\chi^c)^c = - \chi$. 

The presence of the cosmological constant in seven dimensions means that $\zeta$ is not constant, but rather that it satisfies the so-called Killing spinor equation, which for $R_{\rm AdS}=1$ reads
\begin{equation}\label{eq:ksp}
	\nabla_\mu \zeta= \frac12 \gamma^{(7)}_\mu \zeta\ .
\end{equation}
One class of solutions to this equation \cite{breitenlohner-freedman,lu-pope-townsend} is simply of the form
\begin{equation}\label{eq:zeta0}
	\zeta_+ = \rho^{1/2} \zeta^0_+ \ .
\end{equation}
The coordinate $\rho$ appears in (\ref{eq:ads7}), which expresses AdS$_7$ as a warped product of Mink$_6$ and $\mathbb{R}$. $\zeta^0_+$ is a spinor constant along Mink$_6$ and such that $\gamma_{\hat \rho}\zeta^0_+= \zeta^0_+$ (the hat denoting a flat index). 

Just like for Mink$_6\times M_4$, supersymmetry again implies that the norms of the internal spinors $\chi^{1,2}$ should be related to the warping function: $|| \chi_1 ||^2 \pm || \chi_2 ||^2 = c_{\pm} e^{\pm A_3}$, where $c_\pm$ are constant. We will now see, however, that for AdS$_7\times M_3$ actually $c_-=0$. We use the ten-dimensional system in \cite[Eq.~(3.1)]{10d}. As we mentioned in section \ref{sec:pure}, it can be used to derive quickly the system \ref{sub:64}, while applying it directly to AdS$_7\times M_3$ to derive (\ref{eq:73}) is more lengthy. For our purposes, however, it will be enough to apply one equation of that system to the AdS$_7\times M_3$ setup, namely 
\begin{equation}\label{eq:sym}
	d\tilde{K}=\iota_K H
\end{equation}
This is equation (3.1b) in \cite{10d}, but it appeared previously in \cite{hackettjones-smith,figueroaofarrill-hackettjones-moutsopoulos,koerber-martucci-ads}. $K$ and $\tilde{K}$ are the ten-dimensional vector and one-form defined by $K=\frac{1}{64}(\bar{\epsilon}_1 \gamma^{(10)}_M\epsilon_1 +\bar{\epsilon}_2\gamma^{(10)}_M\epsilon_2)dx^{M}$ and $\tilde{K}=\frac{1}{64}(\bar{\epsilon}_1 \gamma^{(10)}_M\epsilon_1 - \bar{\epsilon}_2 \gamma^{(10)}_M\epsilon_2)dx^{M}$. Plugging the decomposed spinors (\ref{eq:eps73}) in these definitions and calling $\beta_1=e^{A_3}(\frac{1}{8}\bar{\zeta}\gamma^{(7)}_{\mu}\zeta) dx^{\mu}$, the part of (\ref{eq:sym}) along AdS$_7$ leads to $e^{A_3}d_7\beta_1(|| \chi_1 ||^2- || \chi_2 ||^2)=(d_7\beta_1)  c_-= 0$, where $d_7$ is the exterior derivative along AdS$_7$. (The right hand side does not contribute, because $H$ has only internal components.) On the other hand, using the Killing spinor equation (\ref{eq:ksp}) in AdS$_7$, we have that $d_7\beta_1= e^{2A_3}(\bar{\zeta}\gamma^{(7)}_{\mu\nu}\zeta) dx^{\mu\nu} \equiv \beta_2$. A spinor in seven dimensions can be in different orbits (defining an SU(3) or an SU(2)$\ltimes \rr^5$ structure \cite{macconamhna,cariglia-macconamhna}), but for none of them the bilinear $\beta_2$ is identically zero. Consequently, the norms of the two Killing spinors have to be equal, namely $c_-=0$.

Let us now see how to translate the spinors $\epsilon_i$ for an AdS$_7 \times M_3$ solution into a language relevant for Mink$_6 \times M_4$. First, we split the seven-dimensional gamma matrices $\gamma^{(7)}_\mu$; the first six give a basis of gamma matrices in six dimensions, $\tilde\gamma^{(6)}_\mu= \rho\gamma^{(7)}_\mu$, $\mu=0,\ldots,5$, while the radial direction, $\gamma^{(7)}_{\hat \rho}=\gamma^{(6)}$ becomes the chiral gamma in six dimensions. (The hat denotes a flat index.) This split is by itself not enough to turn (\ref{eq:gamma73}) into (\ref{eq:gamma64}), because the three-dimensional gamma's in (\ref{eq:gamma73}) have no $\gamma^{(6)}$ in front. This can be cured by applying a change of basis:
\begin{equation}
	\gamma^{(6+4)}_M = O \gamma^{(7+3)}_M O^{-1} \ ,\qquad
	O= \frac1{\sqrt2}(1-i \gamma^{(7+3)}_{\hat \rho}) \ ,
\end{equation}
with, however, a change of basis in six dimensions: $\gamma^{(6)}_\mu \rightarrow -i \gamma^{(6)} \gamma^{(6)}_\mu$. Likewise, the spinors (\ref{eq:eps73}) are related to (\ref{eq:eps64}) by
\begin{equation}\label{eq:Oeps}
	\epsilon_i^{(6+4)}= O \epsilon_i^{(7+3)}\ ,
\end{equation}
if we take 
\begin{equation}\label{eq:etachi}
	\eta_1 = \rho^{1/2}\,\chi_1 \otimes v_+ = \frac1{\sqrt2} \rho^{1/2}\,\chi_1 \otimes {1 \choose -1}
	\ ,\qquad
	\eta_2 = \rho^{1/2}\,\chi_2 \otimes v_\mp = \frac1{\sqrt2} \rho^{1/2}\, \chi_2 \otimes {1 \choose \pm 1}\ .
\end{equation}
Notice that the two $\eta^i$ have equal norm, because the $\chi^i$ have equal norm, as shown earlier. Moreover, since the norm of the $\chi^i$ is $e^{A_3/2}$, and because of the factor $\rho^{1/2}$ in (\ref{eq:etachi}), the $\eta^i$ have norm equal to $\rho^{1/2}e^{A_3/2}$; recalling (\ref{eq:43}), this is equal to $e^{A_4/2}$, as it should. 

Besides (\ref{eq:zeta0}), there is also a second class of solution to the Killing spinor equation $\nabla_\mu \zeta= \frac12 \gamma^{(7)}_\mu \zeta$ on AdS$_7$: it reads $\zeta= (\rho^{-1/2} + \rho^{1/2} x^\mu \gamma^{(7)}_\mu)\zeta^0_-$, where now $\gamma_{\hat \rho}\zeta^0_-= -\zeta^0_-$. If we plug this into (\ref{eq:eps73}) and use the above procedure (\ref{eq:Oeps}) to translate it in the Mink$_6\times M_4$ language, we find a generalization of (\ref{eq:eps64}) where both a positive and negative chirality six-dimensional spinor appear (namely, $x^\mu \gamma_\mu \zeta^0_-$ and $\zeta^0_-$) instead of just a positive chirality spinor $\zeta^0_+$. Because of the $x^\mu \gamma_\mu$ factor, this spinor Ansatz would break Poincar\'e invariance if used by itself; if four supercharges of the form (\ref{eq:eps64}) are preserved, Poincar\'e invariance is present, and these additional supercharges simply signal that an AdS$_7 \times M_3$ solution is ${\cal N}=2$ in terms of Mink$_6\times M_4$.
 

\section{Killing spinors on $S^4$} 
\label{app:kill}

The AdS$_7 \times S^4$ is a familiar Freund--Rubin solution; the flux is taken to be proportional to the internal volume form, $G_4= g \vol_{S^4}$. The eleven-dimensional supersymmetry transformation reads $\left(\nabla_M + \frac1{144}G_{NPQR} (\gamma^{NPQR}{}_M - 8 \gamma^{NPQ} \delta^R_M) \right)\epsilon_{11}=0$; decomposing $\epsilon_{11}= \sum_{a=1}^4\zeta_a \otimes \eta_a + {\rm c.c.}$, and using (\ref{eq:ksp}), one reduces the requirement of supersymmetry (for $R_{\rm AdS}=1$) to taking $g=3/4$, and to the equation 
\begin{equation}\label{eq:kspS4}
	(\nabla_m - \frac12 \gamma \gamma^m) \eta = 0 \ 
\end{equation}
on $S^4$. This is an alternative form of the Killing spinor equation; it was solved in \cite{lu-pope-rahmfeld} in any dimension. However, we are using different coordinates, adapted to the $S^1$ reduction used in section \ref{sub:massless}; we will here solve (\ref{eq:kspS4}) again, using more or less the same method. 

The idea is to start from the easiest components of the equation, and to work one's way to the more complicated ones. Our coordinates in section \ref{sub:massless} are $\alpha$, $\beta$, $\gamma$, $y$, the latter being the reduction coordinate. Our vielbein reads $e^1= d \alpha$, $e^2=\frac12 \sin(\alpha) d \beta$, $e^3= \frac12 \sin(\alpha) \sin(\beta) d \gamma$, $e^4= \frac12 \sin(\alpha) (dy + \cos(\beta) d \gamma)$. We begin with the $\alpha$ component of (\ref{eq:kspS4}):
\begin{equation}
	\del_\alpha \eta = \frac12 \gamma \gamma_1 \eta \qquad \Rightarrow \qquad 
	\eta= e^{\frac12 \alpha \gamma \gamma_1} \eta_1 \ .
\end{equation}
The next component we use is 
\begin{equation}
	\left(\del_\beta - \frac14 \cos(\alpha)\right) \eta = \frac14 \sin(\alpha) \gamma \gamma_2 \eta \ .
\end{equation}
This can be manipulated as follows:
\begin{equation}
	0=\left( \del_\beta - \frac14 e^{\alpha \gamma \gamma_1} \gamma_{12}\right) \eta = e^{\frac12\alpha \gamma \gamma_1} \left( \del_\beta - \frac14 \gamma_{12}\right) \eta_1 \qquad \Rightarrow \qquad 
	\eta_1 = e^{\frac14 \beta \gamma_{12}} \eta_2 \ .
\end{equation}
We proceed in a similar way for the two remaining coordinates; the details are complicated, and we omit them here. The final result is 
\begin{equation}
	\eta = \exp\left[\frac \alpha2 \gamma \gamma_1\right] \exp\left[\frac \beta 4 \gamma_{12} + \frac{\beta- \pi}4 \gamma_{34}\right]
	\exp\left[\frac{y+\gamma}4 \gamma_{13} + \frac{y-\gamma}4 \gamma_{24}\right] \eta_0
\end{equation}
where $\eta_0$ is a constant spinor. When we reduce, we demand that $\del_y \eta=0$, which becomes $(\gamma_{13}+ \gamma_{24}) \eta_0 = 0 $; this condition indeed keeps two out of four spinors, as anticipated in our discussion in section \ref{sub:massless}.

\section{Sufficiency of the system (\ref{eq:73})} 
\label{app:suff}

In section \ref{sub:73} we obtained the system of equations (\ref{eq:73}) starting from (\ref{eq:64}) and using the fact that AdS$_7$ can be considered as a warped product of Mink$_6$ and $\rr$. In this section we will explain how one can show that (\ref{eq:73}) is completely equivalent to supersymmetry for $\mathrm{AdS}_7 \times M_3$ with a direct computation. Our strategy will be very similar to the one in \cite[Sec.~A.4]{gmpt3}, with some relevant differences that we will promptly point out.

To begin with, we write the system of equations resulting from setting to zero the type II supersymmetry variations (of gravitinos and dilatinos) using the spinorial decomposition (\ref{eq:eps73}):\footnote{We choose to show the equivalence in the IIA case, hence we pick $\epsilon_1^{(7+3)}$ and $\epsilon_2^{(7+3)}$ with opposite chirality.}
\begin{subequations} \label{eq:SUSYvariation73m}
\begin{align}
  & \left( D_m - \frac 14 H_m \right) \chi_1 -\frac{ e^\phi}{8} F \sigma_m \chi_2 = 0 \ , \label{eq:susy73m1}\\
  & \left( D_m + \frac 14 H_m \right) \chi_2 -\frac{ e^\phi}{8} \lambda(F) \sigma_m \chi_1 = 0 \ , \label{eq:susy73m2}\\
  & \frac 12 e^{-A} \chi_1 - \frac{i}{2} \partial A \chi_1 + i \frac{e^\phi}{8} F \chi_2 = 0 \ , \\
  & \frac 12 e^{-A} \chi_2 + \frac{i}{2} \partial A \chi_2 - i \frac{e^\phi}{8} \lambda(F) \chi_1 = 0 \ , \\
  & \left(D - \frac 14 H \right)\chi_1 + i \frac 72 e^{-A} \chi_1 + \left(\frac 72 \partial A - \partial \phi \right) \chi_1 = 0 \ , \\
  & \left(D + \frac 14 H \right)\chi_2 - i \frac 72 e^{-A} \chi_2 + \left(\frac 72 \partial A - \partial \phi \right) \chi_2 = 0 \ .
\end{align}
\end{subequations}
As in \cite[Sec.~A.4]{gmpt3}, we introduce a set of intrinsic torsions $p_m^a$, $q_m^a$, and $T^a$, $\hat{T}^a$, with $a=1,2$:
\begin{subequations} \label{eq:intrinsictorsm}
\begin{align}
& \left(D_m - \frac 14 H_m \right) \chi_1 \equiv p^1_m \chi_1 + q^1_m \chi_1^c \ , \qquad \left(D_m + \frac 14 H_m \right) \chi_2 \equiv p^2_m \chi_2 + q^2_m \chi_2^c \ , \label{eq:intrtors1}\\
& \left(D- \frac 14 H \right) \chi_1 \equiv  T^1 \chi_1 + \hat{T}^1 \chi_1^c \ , \qquad \left(D + \frac 14 H \right) \chi_2 \equiv  T^2 \chi_2 + \hat{T}^2 \chi_2^c \ , \label{eq:intrtors2}
\end{align}
\end{subequations}
where $D=\gamma_{(7)}^m D_m$, $H_m \equiv \frac12 H_{mnp}\gamma_{(7)}^{np}$, $H \equiv \frac 16 H_{mnp} \gamma_{(7)}^{mnp}$ as usual. We used the fact that $\chi_1$ and $\chi_1^c$ (or $\chi_2$ and $\chi_2^c$) constitute a basis for the three-dimensional spinors. Taking tensor products of these two bases, we also obtain a basis for bispinors, on which we can now expand $F$:
\begin{equation} \label{eq:fluxbispinorsm}
F \equiv R_{00}\, \chi_1 \otimes \chi_2^\dagger + R_{10} \,\chi_1^c \otimes \chi_2^\dagger + R_{01} \,\chi_1 \otimes \chi_2^{c \dagger} + R_{11}\, \chi_1^c \otimes \chi_2^{c \dagger} \ .
\end{equation}
Using (\ref{eq:intrinsictorsm}) and (\ref{eq:fluxbispinorsm}) in (\ref{eq:SUSYvariation73m}), we can rewrite the conditions for unbroken supersymmetry as a set of equations relating the intrinsic torsions to the coefficients $R_{ij}$. Let us call this system of equations the ``spinorial system''. Using instead (\ref{eq:intrinsictorsm}) and (\ref{eq:fluxbispinorsm}) in (\ref{eq:73}), we obtain a second set of equations, again in terms of the intrinsic torsions and $R_{ij}$; let us call this system the ``form system''. Our aim is to show the equivalence between the spinorial and the form systems.

Although we are using the same technique appearing in \cite[Sec.~A.4]{gmpt3} (there applied to four-dimensional vacua), proving this equivalence in the case at hand is more involved. Relying on a superficial counting, it would seem that the form system contains fewer equations than the spinorial one. To see why this happens, we first notice that the definitions (\ref{eq:intrinsictorsm}) are redundant. Indeed the torsions $T^a$ and $\hat{T}^a$ can be rewritten in terms of the torsions $p^a$, $q^a$ and $H$; however, in three dimensions, $\gamma^{(7)}_{mnp}$, hence $H$, is proportional to the identity (use (\ref{eq:sla*}) with $\alpha=H$). Thus in (\ref{eq:intrtors2}) four complex numbers ($T$'s and $\hat{T}$'s) are used to describe a single real number $H$. This suggests that some of the equations in the spinorial system are redundant and could be dropped. However, this redundancy is not manifest.

To make it manifest, we could use the following strategy. On the one hand (\ref{eq:susy73m1}) and (\ref{eq:susy73m2}) give a natural expansion of the torsions $p^a$ and $q^a$ in terms of the vielbein $e^b$, with $a \neq b$, defined by the spinor $\chi_b$ (see (\ref{eq:ccd}) and (\ref{eq:ccb})); that is, they transform into equations for the components $q^1 \cdot e_3^2$, $q^1 \cdot e_1^2$ and so forth. On the other hand the intrinsic torsions $T^a$ and $\hat{T}^a$ give expressions like $q^1 \cdot e_3^1$, $q^1 \cdot e_1^1$. Therefore, we would need a formula relating the vielbein $e^1$ defined by $\chi_1$ to the vielbein $e^2$ defined by $\chi_2$.

Actually, there exists a simpler method. Indeed we can use the following equations,
\begin{equation}
\label{eq:redundantm}
\begin{split}
  & d_H (e^{2A_3 - \phi} \mathrm{Re} \psi^1_-) = 0 \ , \\
  & d_H (e^{4A_3 - \phi} \mathrm{Im} \psi^1_-) = 0 \ , \\
  & d_H (e^{4A_3 - \phi} \psi^2_-) = 0 \ ,
\end{split}
\end{equation}
obtained by simply applying $d_H$ to the equations (\ref{eq:73I1}), (\ref{eq:73R1}) and (\ref{eq:732}) respectively (in other words, they are redundant with respect to the original system (\ref{eq:73})). If we now express (\ref{eq:redundantm}) in terms of (\ref{eq:intrinsictorsm}), and add the resulting equations to the form system we obtained earlier, we obtain a new, equivalent expression for the form system. With some linear manipulations, it can now be shown that it is equivalent to the spinorial system. This concludes our alternative proof that (\ref{eq:73}) is completely equivalent to the requirement of unbroken supersymmetry.




\providecommand{\href}[2]{#2}

\end{document}